\documentclass[aps,twocolumn,floatfix,altaffilletter,superscriptaddress,preprintnumbers,tightenlines,showpacs,showkeys,notitlepage,nofootinbib]{revtex4-2}

\usepackage[T1]{fontenc}
\usepackage[table]{xcolor}
\usepackage[colorlinks=true,citecolor=blue,linkcolor=blue]{hyperref}
\usepackage{braket,graphicx,physics,amsthm,amssymb,amsfonts,booktabs,siunitx,mathrsfs}
\usepackage{mathtools}
\usepackage{slashed}
\usepackage[capitalise, english]{cleveref}
\usepackage[normalem]{ulem}
\usepackage{gensymb}
\usepackage{nicematrix}

\allowdisplaybreaks  

\makeatletter

    \def\CT@@do@color{%
      \global\let\CT@do@color\relax
            \@tempdima\wd\z@
            \advance\@tempdima\@tempdimb
            \advance\@tempdima\@tempdimc
    \advance\@tempdimb\tabcolsep
    \advance\@tempdimc\tabcolsep
    \advance\@tempdima2\tabcolsep
            \kern-\@tempdimb
            \leaders\vrule
                    \hskip\@tempdima\@plus  1fill
            \kern-\@tempdimc
            \hskip-\wd\z@ \@plus -1fill }
    \makeatother

\newcommand{\beq}{\begin{equation} }
\newcommand{\eeq}{\end{equation}} 
\newcommand{\bi}{\begin{itemize} }
\newcommand{\ei}{\end{itemize} }

\makeatletter
\providecommand*{\diff}%
	{\@ifnextchar^{\DIfF}{\DIfF^{}}}
\def\DIfF^#1{%
	\mathop{\mathrm{\mathstrut d}}%
		\nolimits^{#1}\gobblespace}
\def\gobblespace{%
	\futurelet\diffarg\opspace}
\def\opspace{%
	\let\DiffSpace\!%
	\ifx\diffarg(%
		\let\DiffSpace\relax
	\else
		\ifx\diffarg[%
			\let\DiffSpace\relax
		\else
  			\ifx\diffarg\{%
				\let\DiffSpace\relax
			\fi\fi\fi\DiffSpace}

\definecolor{Red}{rgb}{1.,0.,0.}

\usepackage{xspace}

\definecolor{cborange}{HTML}{e69f00}
\definecolor{cbgreen}{HTML}{009e73}
\definecolor{cbyellow}{HTML}{f1dd42}
\definecolor{cblblue}{HTML}{56b4e9}
\definecolor{cbblue}{HTML}{0072b2}
\definecolor{defgrey}{HTML}{9f9f9f}
\definecolor{defgreen}{HTML}{8eba42}

\newcommand{\Luv}{{\Lambda_\mathrm{UV}}}

\begin{document}

\preprint{}

\title{Invisible jets from composite neutrinos}

\author{Matteo Borrello}
\affiliation{INFN, Sezione di Firenze Via G. Sansone 1, 50019 Sesto Fiorentino, Italy and\\ Department of Physics and Astronomy, University of Florence, Italy}
\author{Marco Costa}
\affiliation{Perimeter Institute for Theoretical Physics, 31 Caroline St N, Waterloo, ON N2L 2Y5, Canada}
\author{Diego Redigolo}
\affiliation{INFN, Sezione di Firenze Via G. Sansone 1, 50019 Sesto Fiorentino, Italy and\\ Department of Physics and Astronomy, University of Florence, Italy}

\begin{abstract}
We propose a novel experimental probe of neutrino couplings to a composite sterile sector, leveraging the unique signature of neutrino disintegration into ``invisible jets'' in high-energy neutrino scattering. Focusing on scenarios where the invisible jet invariant mass significantly exceeds the confinement scale, we compute production rates within the conformal window. In this regime, invisible jet production leads to an energy-dependent enhancement of the neutral-to-charged current ratio in neutrino–nucleus scattering. Using NuTeV measurements, we derive new bounds and assess the sensitivity of upcoming experiments such as SHiP and the Forward Physics Facility at CERN. We sketch models where these probes surpass constraints from electroweak gauge boson and Higgs invisible branching ratios. In contrast, neutrino–electron scattering modifications are suppressed by the lower center-of-mass energy and are unlikely to be observable at DUNE.
\end{abstract}


\maketitle

\section{Introduction}

Neutrino masses signal physics beyond the Standard Model (SM), but their origin remains uncertain. The standard seesaw mechanism generates Majorana masses via the coupling $\Delta\mathcal{L}_N = y_N H L N$, yielding $m_\nu \sim U_{N\nu}^2 M_N$~\cite{Minkowski:1977sc, Mohapatra:1979ia, Foot:1988aq, Gelmini:1980re}, where $U_{N\nu} = y_N v / M_N$ is the active-sterile mixing and $v = 246\,\text{GeV}$ the electroweak (EW) vacuum expectation value (VEV). Current limits on neutrino masses~\cite{Lesgourgues:2013sjj, Lattanzi:2017ubx, Pascoli:2019wpp, Hernandez:2016kgx} require sterile neutrinos to be either very heavy or weakly coupled, suppressing mixing and hindering detection.

The inverse seesaw~\cite{Mohapatra:1986bd, Gonzalez-Garcia:1988okv} offers a testable alternative, introducing pseudo-Dirac sterile states with large Dirac mass $M_N$ and small lepton-number violating Majorana mass $\mu_N$. In this framework, $m_\nu \sim U_{N\nu}^2 \mu_N$, allowing for sizable mixing and enhancing prospects for experimental searches of sterile neutrinos.

Motivated by the inverse seesaw, considerable attention has been given to the signatures of weakly coupled sterile neutrinos~\cite{Dasgupta:2021ies}, while scenarios involving strongly coupled sterile sectors interacting with SM neutrinos through higher-dimensional operators remain less explored~\cite{Arkani-Hamed:1998wff, vonGersdorff:2008is, Grossman:2010iq, Chacko:2020zze, Ahmed:2023vdb}. In this letter, we propose a novel and distinctive signature of neutrinos interacting with a composite sector, potentially observable in high-energy neutrino experiments. 
In our scenario, SM neutrinos scatter off electrons or nucleons via neutral current interactions, producing an invisible jet—a collimated spray of hidden-sector particles. We focus on the experimentally challenging regime where the invisible jet constituents are either long-lived on detector scales—typical of highly irrelevant portals—or neutral under the SM, as expected in models where the sterile neutrino sector also accounts for the dark matter relic abundance~\cite{Ahmed:2023vdb, Hong:2024zsn}.

The presence of an invisible jet would manifest as an energy-dependent enhanced neutral-current event rate. This is constrained by previous high-energy neutrino experiments such NuTeV~\cite{NuTeV:2001whx} and can be further probed by future experiments such as DUNE~\cite{DUNE:2020lwj}, SHiP~\cite{SHiP:2021nfo}, and FASER$\nu$~\cite{FASER:2020gpr} especially in its high-luminosity upgrade, the planned Forward Physics Facility (FPF) at the LHC \cite{MammenAbraham:2024gun,Feng:2022inv}.  We discuss to what extent neutrino disintegration into an invisible jet can be distinguished from standard electron–neutrino scattering (see Sec.~\ref{sec:escatt}) and nucleon–neutrino deep-inelastic scattering (see Sec.~\ref{sec:Nscatt}). 

The unique features of this signal open a new experimental avenue for probing composite neutrino interactions, while presenting novel challenges for optimizing signal–background discrimination in next-generation neutrino experiments.

\section{Setup and Results}\label{sec:setup}
\begin{figure}[t!]
     \centering
     \includegraphics[width=1\linewidth]{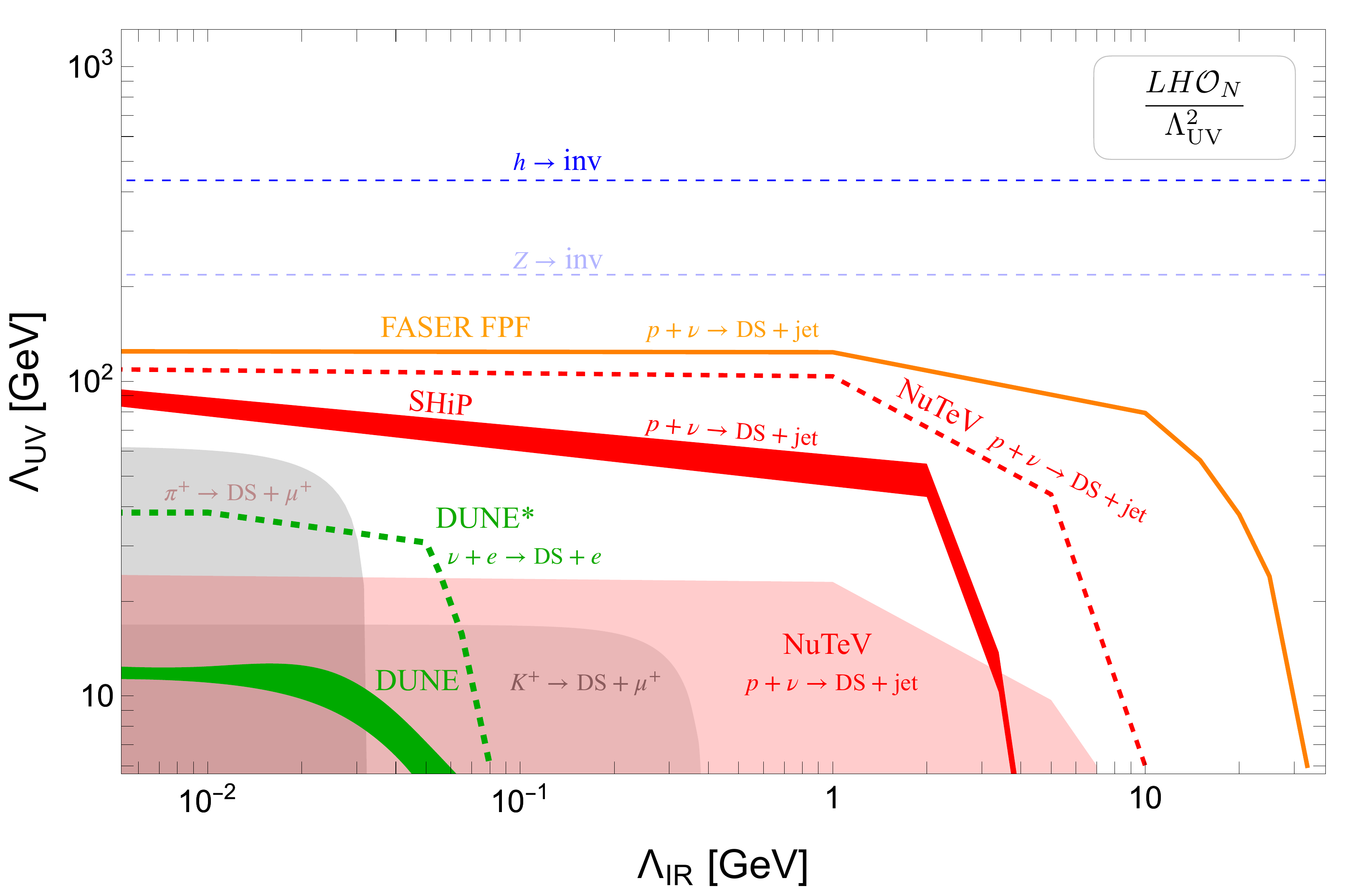}
     \caption{Plot of the portal in Eq.~\eqref{eq:compositeN} for \(\Delta_N = 7/2\), \(c_N = 100\) (\(A_N = 3.27\)). \textbf{Red-shaded}: Excluded by NuTeV~\cite{NuTeV:2001whx}. \textbf{Red band}: SHiP sensitivity with $Q^2_{\rm{SM}}>2\,\rm{GeV}$ (lower line) and with all the events (upper line) (see Sec.~\ref{sec:Nscatt}) \textbf{Orange:} Forward Physics Facility sensitivity. \textbf{Green band}: DUNE sensitivity; stronger with tau-optimized flux, \textbf{dashed} for narrower flux $\delta E_\nu/E_\nu=20\%$. (see Sec.~\ref{sec:escatt}) \textbf{Dark Blue}: Higgs invisible branching ratio constraints. \textbf{Light Blue}: Z invisible branching ratio constraints.  \textbf{Gray-shaded}: Excluded by meson decays (App.~\ref{app:meson})}.
     \label{fig:moneyplot}
 \end{figure}

 We follow Ref.~\cite{Chacko:2020zze} and model the sterile sector as a strongly coupled theory interacting with muon neutrinos via  
\begin{equation}
\Delta\mathcal{L}_N = \frac{y_\mu H L_\mu \mathcal{O}_N}{\Lambda_{\rm{UV}}^{\Delta_N - 3/2}}, \label{eq:compositeN}
\end{equation}
where \(\mathcal{O}_N\) is a fermionic operator of dimension \(\Delta_N\), and \(\Lambda_{\rm UV}\) is the UV scale. This  is just an example of a much wider array of models that can be tested through the idea proposed here (see for example Appendix~\ref{app:dipole} for a different model).

The sterile sector has a mass gap \(\Lambda_{\rm IR} \ll \Lambda_{\rm UV}\) and admits a description in terms of single particle states at \(E \sim \Lambda_{\rm IR}\). Their lifetime scales as  
\begin{equation}
\tau_{N}\approx\!  1.6\cdot10^{-10+4\Delta_N}\,\mathrm{km} \left[\frac{10\,\mathrm{MeV}}{\Lambda_{\rm{IR}}}\right]^{2\Delta_N}\! \left[\frac{\Lambda_{\rm{UV}}}{1\,\mathrm{GeV}}\right]^{2\Delta_N - 3},
\label{eq:lifetime}
\end{equation}
yielding \(\tau_N \sim  10^4\,\mathrm{km}\) for \(\Delta_N = 7/2\). We focus on $\Delta_N=7/2$ hereafter.\footnote{Lepton number violation is parameterized by a scalar operator in the sterile sector, $\Delta \mathcal{L}_{\slashed{L}} = \epsilon \mathcal{O}_{\slashed{L}} / \Lambda_{\rm{UV}}^{\Delta_{\slashed{L}} - 4}$. As in the inverse seesaw, small neutrino masses are technically natural and can be made arbitrary light for $\epsilon \ll 1$. The masses arise from the three-point function $\langle \mathcal{O}_N \mathcal{O}_N \mathcal{O}_{\slashed{L}} \rangle$, which induces the Weinberg operator $(HL)^2$. Unlike Ref.~\cite{Chacko:2020zze}, we do not construct a full model for neutrino masses, but note that for $\Delta_N = 7/2$ an irreducible UV contribution to $(HL)^2$ is expected.}
This leads to long-lived invisible jets that evade direct detection but leave their unique imprints in the nature of the disintegration process.

Disintegration proceeds dominantly via Z-boson exchange neutral current interactions, with negligible Higgs contribution. As can be seen by the diagrams in Fig.~\ref{fig:el} and Fig.~\ref{fig:N} it is crucial to access the dark sector continuum to propagate a virtual active neutrino line. For \(\Delta_N \geq 7/2\), the disintegration cross section is dominated by the high-energy continuum of unparticle states~\cite{Georgi:2007ek,Georgi:2007si} and can be reliably computed within the conformal window as long as \(\Lambda_{\rm IR} \lesssim \sqrt{s} \lesssim \Lambda_{\rm UV}\), independently of details of the composite spectrum~\cite{Contino:2020tix}. In this regime, the inclusive cross section relates via the optical theorem to the imaginary part of the conformal two-point function,  
\begin{equation}
{\rm Im}[i\langle T[\mathcal{O}_N(p) \bar{\mathcal{O}}_N(-p)]\rangle] = A_\Delta p^{2\Delta_N - 5} \slashed{p},
\label{eq:2point}
\end{equation}
with $A_\Delta = \frac{2^{3 - 2\Delta_N} \pi c_N}{\Gamma(\Delta_N + 1/2) \Gamma(\Delta_N - 3/2)},$ and \(c_N > 0\) the central charge of the dark sector. While the imaginary part is always regular, Appendix~\ref{eq:CFT} clarifies the treatment of UV divergences in the full correlator.  

Table~\ref{tab:exp} summarizes the neutrino experiments considered. Our main results appear in Fig.~\ref{fig:moneyplot}, NuTeV conservative limits are obtained requiring the signal to be less than the measured data in Ref.~\cite{NuTeV:2001whx} while the expected sensitivities of NuTeV, SHiP and FPF compare the signal to the measured (for NuTeV) or expected (for SHiP and FPF)  background uncertainties. DUNE projections for electron recoils exploit the full kinematics of the event and are discussed in Sec.~\ref{sec:escatt}. 

Signal and background event counts scale as
$S,B = \sigma_{S,B} \times \sigma^{-1}_{\rm target} \times \mathcal{L}_{\rm exp},$
with $\sigma_S \sim \frac{s^2}{v^2 \Lambda_{\rm UV}^4} \left(\frac{s}{\Lambda_{\rm UV}^2}\right)^{\Delta_N - 7/2}$ and  $\sigma_B \sim \frac{s}{v^4}$.
For \(\Delta_N > 5/2\), the signal grows faster with energy than the SM background, favoring nuclear recoils (\(s_p = 2m_p E_\nu\)) over electron recoils (\(s_e =2 m_e E_\nu\)) and high-energy neutrino beams such as NuTeV, SHiP and FPF. We focus on DIS events calculable within the conformal window, excluding DUNE—where quasi-elastic scattering dominates~\cite{Formaggio:2012cpf} and FASER\(\nu\) with the current luminosity where our EFT description breaks down.

The signal explored in this work can be parametrized as an energy-dependent correction to the ratio of neutral current (NC) to charged current (CC) neutrino-nucleon scattering. Generalizing the definition from Ref.~\cite{Davidson:2003ha}, we define
\begin{equation}\label{eq:R_ratio_def}
    R_\nu(s) \equiv \frac{\sigma(\nu\,N \to \nu\,X)}{\sigma(\nu\,N \to \mu^-\,X)} = g_L^2 + r\,g_R^2 + \Delta g^2(s)\,,
\end{equation}
where \(g_L^2 = \tfrac{1}{2} - \sin^2\theta_W + \tfrac{5}{9}\sin^4\theta_W\) and \(g_R^2 = \tfrac{5}{9}\sin^4\theta_W\) are the effective left- and right-handed isoscalar couplings, with $\sin\theta_{\rm w}^2\simeq0.22$ and \(r = \sigma(\bar\nu\,N \to \mu^+ X)/\sigma(\nu\,N \to \mu^-\,X) \approx 0.5\) is the ratio of antineutrino to neutrino CC cross sections. The term \(\Delta g^2(s)\) encodes potential new physics contributions arising from a portal operator of dimension \(\Delta_N\).

In the limit of a narrow neutrino beam, where the energy spread \(\Delta_s\) is much smaller than the average squared center-of-mass energy \(\bar{s}\), the ratio of new physics neutral current events over the charged current ones  scales as
\begin{equation}
    \frac{\rm{BSM}\,\rm{NC}'s}{\rm{BSM}\,\rm{CC}'s} \sim \frac{\bar{s}\, v^2}{\Lambda_{\text{UV}}^4} \left( \frac{\bar{s}}{\Lambda_{\text{UV}}^2} \right)^{\Delta_N - \frac{7}{2}}\,.\label{eq:scalingratio}
\end{equation}
This expression generalizes the class of BSM searches based on deviations in the neutral-to-charged current ratio. The strong dependence on the mean beam energy \(\bar{s}\) suggests that varying the neutrino beam energy configuration could enhance sensitivity to such new physics and help disentangle its effects from SM expectations.

In contrast, for a broad neutrino energy fluxes—such as that of the NuTeV experiment—where \(\Delta_s \gg \bar{s}\), the correction becomes effectively independent of the beam energy and the constant shift can be obtained at the leading order in $\mathcal{O}(\bar{s}/\Delta_s)$ by replacing $\bar{s}\to \Delta_s$ in Eq.~\eqref{eq:scalingratio}. As a consistency check of our results, we used this constant shift to reinterpret the NuTeV sensitivity by translating the experimental bound on \(g_L^2\) reported in Ref.~\cite{Davidson:2003ha} into a constraint on $\Delta g^2(s)$.

We impose a lower bound \(\Lambda_{\rm IR} \gtrsim 10\,\mathrm{MeV}\) from BBN constraints~\cite{Chacko:2020zze}. Rare meson decays \(M^+ \to \mu^+ + \mathrm{inv.}\) (\(M^+ = \pi^+, K^+\)) computed in Appendix~\ref{app:meson} provide additional limits extracted from the current uncertainties in the SM branching ratios~\cite{ParticleDataGroup:2020ssz}. The broad missing mass distribution of invisible jets weakens these constraints compared to standard sterile neutrino searches.

As shown in Fig.~\ref{fig:moneyplot}, invisible EW gauge bosons and expecially Higgs decays place stringent constraints on dark sectors assuming the conformal window extends up to \(m_h\).  Instead, these bounds can be relaxed if the conformal window breaks at an intermediate scale \(\Lambda_\star < m_h\), where the theory becomes UV free.\footnote{UV-free gauge theories with matter can have conformal windows---parameter regions where the theory flows to a nontrivial IR conformal fixed point. In QCD-like theories with \(N_* < N_f < \frac{11}{2} N_c\), the upper window hosts a weakly coupled fixed point accessible perturbatively, while the lower end (defined by $N_*$) may feature strongly coupled fixed points whose existence is less certain~\cite{Kaplan:2009kr,Jarvinen:2011qe,Alvares:2012kr,DiPietro:2020jne}. Similarly, in \(\mathcal{N}=1\) SUSY QCD with \(\frac{3}{2} N_c < N_f \ll 3 N_c\), strongly coupled IR fixed points can be studied via Seiberg dualities~\cite{Intriligator:1995au}. In our setup, the IR fixed point is perturbed by a marginally relevant operator, driving confinement and stabilizing the hierarchy between \(\Lambda_{\rm UV}\) and \(\Lambda_{\rm IR}\).}

As an example, one can consider a UV completion where the Higgs couples directly to a heavy neutral lepton \(\Psi\) with mass $\Lambda_\star < M_\Psi < m_h.$ This fermion has a sizable invisible decay width into dark states, but the Higgs only ``sees'' this weakly coupled state rather than the full strongly coupled conformal sector. Below \(M_\Psi\), integrating out \(\Psi\) while the dark sector is still perturbative generates the effective portal which subsequently flows into the conformal window, reproducing the neutrino disintegration phenomenology discussed above.

A sketch of an explicit UV completion is given by 
\begin{equation}\label{eq:uvmodel}
\mathcal{L}_{\rm UV} = \tilde{y}_\mu H L_\mu \Psi + M_\Psi \bar{\Psi} \Psi + \frac{\bar{\Psi} \lambda \Phi^2}{\Lambda_N},
\end{equation}
where \(\Psi\) is a singlet fermion under both the SM and dark gauge groups, and \(\lambda, \Phi\) are dark sector fields charged under the dark gauge group. 

On the one hand, the Higgs invisible decay constraint on \(h \to \nu \Psi\) requires \(\tilde{y}_\mu \lesssim 10^{-2}\)~\cite{Bernal:2023coo}, translating into
$\sqrt{\Lambda_N M_\Psi} \lesssim 0.1\, \Lambda_{\rm UV}$
for \(y_\mu \sim \mathcal{O}(1)\). On the other hand, perturbativity demands
$\Lambda_N \gtrsim \frac{\Lambda_{\rm UV}}{16 \pi^2}$,
leaving still viable parameter space where both the Higgs constraints are satisfied and the neutrino disintegration signal remains abundant.

This simple example illustrates how experimental probes at center of mass energies larger than the ones of neutrino experiments are highly sensitive to the details of the UV completion of the sterile sector. For this reason, we do not consider LHC bounds or constraints from precision electroweak observables in our analysis. The latter depend on UV-sensitive counterterms such as $\bar{\ell} \Box\gamma^\mu\partial_\mu \ell$ that lie beyond the reach of our EFT.

We note that such counterterms, when contracted with quark or electron $Z$ currents as in $ \bar{\nu}\gamma_\mu \Box \nu \bar q\gamma^{\mu} q$, can generate non-renormalizable operators that mimic the energy dependence coming from our BSM signal. Notice that also operators containing other combinations of fermions and derivatives are expected, leading to a UV dependent contribution to the number of neutral current events with an undetermined sign and a behavior $\sim s^2$ with energy, but with the kinematics of SM events.

\begin{table}[t!]

    \centering
    
    \begin{tabular}{c|cccc}
                Exp. & $N_\nu$ & $\langle E_\nu\rangle$ [GeV]&$\sigma_E^2$ [GeV$^2$]&$\sigma^{-1}_\text{target}$ [GeV$^2$]\\
        \hline
   \rowcolor{green!20}   DUNE   & $10^{20}$   & 3& 4 &0.02  \\
       
  \rowcolor{green!20}     DUNE$^*$  & $10^{20}$   & 20& 8 &0.02  \\
       \hline
   \rowcolor{pink!20}     NuTeV  & $ 10^{17}$  & 60&$10^3$&0.02\\
        \hline
     \rowcolor{red!20}   SHiP   & $5\times 10^{16}$   & 40&32 &0.2    \\
        \hline
    \rowcolor{orange!20}    FASER$\nu$  & $ 10^{12}$  & 300&$10^5$&0.2  \\
     \rowcolor{orange!20}    FASER-FPF  & $ 10^{15}$  & 300&$10^5$&0.2      
    \end{tabular}
    \caption{Fluxes and luminosities for the different neutrino experiments assumed here. The squared width $\sigma_E^2$ is the order of magnitude of the width of the realistic fluxes plotted in See Appendix~\ref{app:eNscat} for simplicity. For SHiP we assumed a gaussian flux with variance $\sigma_E^2$. The luminosity at FASER-FPF is chosen to have $10^6$ CC events from $\nu_\mu$. }

    \label{tab:exp}
    \end{table}

\section{Scattering against electron}\label{sec:escatt}
\begin{figure}[t!]
     \centering
     \includegraphics[width=1\linewidth]{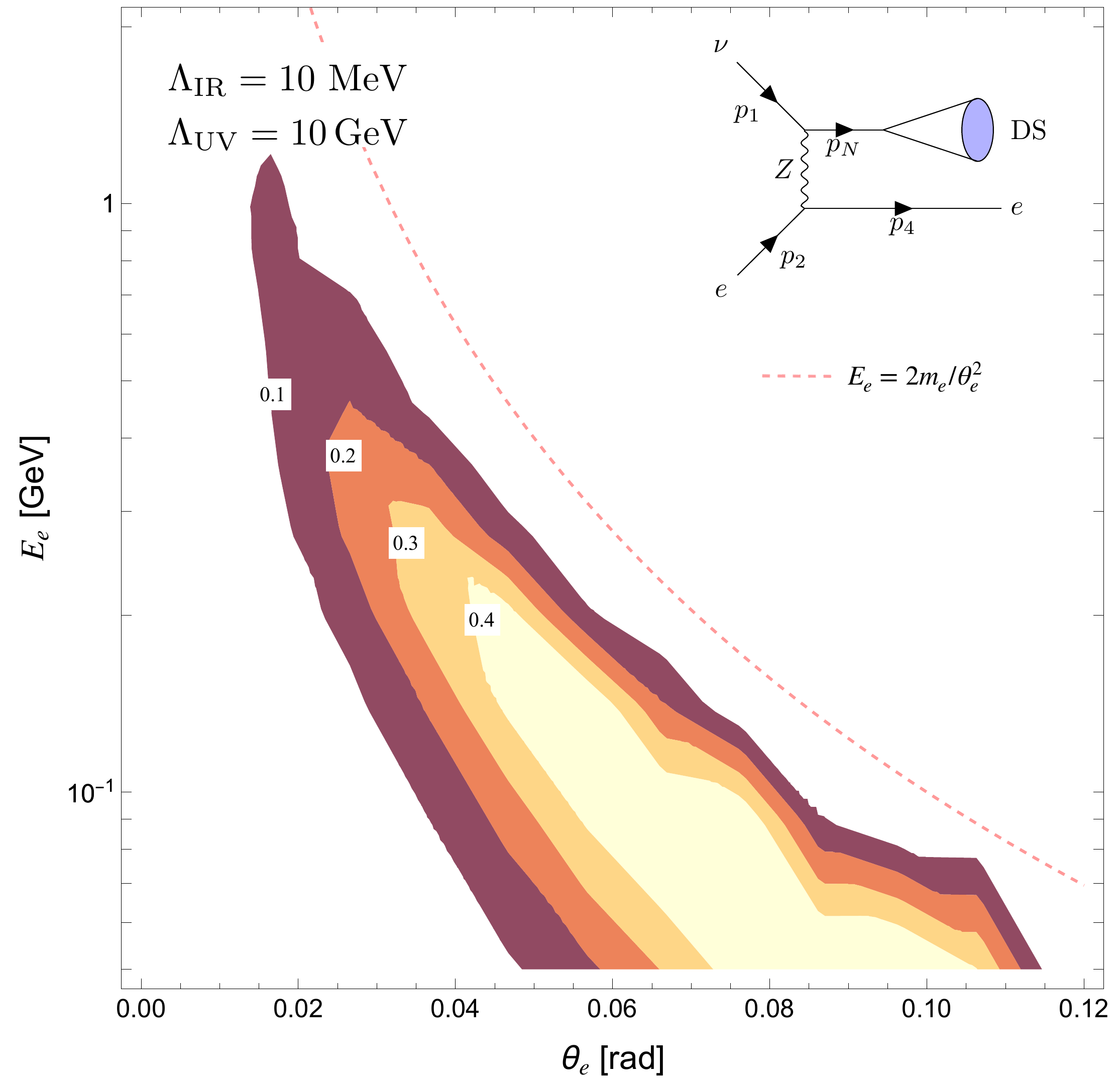}
     \caption{Log-likelihood for electron recoil events at DUNE for the DUNE baseline neutrino flux. The red dashed line indicates the boundary of the forward region demonstrating that the whole likelihood is concentrated in the forward region (see Appendix~\ref{sec:electron}).}
     \label{fig:el}
 \end{figure}
 The electron recoil case is simple enough to clearly illustrate the key features of our signal and the strategy to distinguish it from background. Defining $p_N^2=m_N^2$ to be the invariant mass of the invisible jet the differential signal cross section for generic $\Delta_N$ can be written as 
 \begin{equation}
 \frac{d\sigma_L}{d m_N^2}=\frac{A_N l_e^2 y_\mu^2}{96\pi^2} \frac{s_e}{v^2\Lambda_{\rm UV}^4}\left[\frac{m_N^2}{\Lambda^2_{\rm{UV}}}\right]^{\Delta_N-7/2}\left[1-\frac{m_N^2}{s_e}\right]^2\label{eq:diffe}
 \end{equation}
 for the left-handed EW neutral current with $l_e=1/2-\sin\theta_{\rm w}^2$. The contribution from the right-handed EW current is obtained by replacing $l_e\to \frac{r_e}{6}\left(2+\frac{m_N^2}{s_e}\right)$ with $r_e=\sin\theta_{\rm w}^2$ and should be summed to the left one above to obtain the total cross section. 
 
As long as \( m_N^2 \ll s_e \) and invisible jet production is kinematically allowed, the differential cross section exhibits a power-law dependence on \( m_N^2 \): it increases for \( \Delta_N > 7/2 \), decreases for \( \Delta_N < 7/2 \), and remains constant at \( \Delta_N = 7/2 \). This behavior highlights the concept of UV vs. IR dominance: for \( \Delta_N \geq 7/2 \), the ultraviolet (high-mass) region dominates the total cross section, whereas for \( \Delta_N < 7/2 \), the infrared (low-mass) region is dominant. The requirement of UV dominance ensures that the signal is both calculable—via Eq.~\eqref{eq:2point}—within the conformal window and clearly distinguishable from the background, which arises from neutrino–electron elastic scattering.

For a monochromatic neutrino beam, the elastic scattering cross section \cite{Formaggio:2012cpf} peaks at \( m_N^2 = 0 \), corresponding to the invariant mass of the outgoing neutrino. In this idealized (though unrealistic) scenario, the signal is effectively background-free, allowing a straightforward sensitivity estimate. Integrating Eq.~\eqref{eq:diffe} over \( m_N^2 \in [4\Lambda_{\rm{IR}}^2, s_e] \), and requiring at least two signal events at DUNE, yields a reach of \( \Lambda_{\rm{UV}} \simeq 36\,\mathrm{GeV} \) for \( \Lambda_{\rm{IR}} \ll \sqrt{s_e} \), with \( \sqrt{s_e} \simeq 0.14\,\mathrm{GeV} \) and a total neutrino flux of \( N_\nu =  3\times 10^{20} \). Remarkably, this back-of-the-envelope estimate aligns well with the reach obtained from a one-dimensional log-likelihood analysis for a hypothetical narrow-band neutrino beam at DUNE, as shown by the dashed line in Fig.~\ref{fig:moneyplot}.

For a realistic, energy-broad neutrino beam, the incoming flux smears the missing mass distribution, requiring a more refined analysis to distinguish the signal. We implement a binned log-likelihood in electron energy \( E_e \) and angle \( \theta_e \), $\lambda\approx \sum_i S_i^2/B_i$ where \( S_i \) and \( B_i \) are the signal and background in the \( i \)-th bin. Setting \( \lambda = 2.71 \) yields the 95\% CL exclusion region in the \( (\Lambda_{\rm IR}, \Lambda_{\rm UV}) \) plane, shown in Fig.~\ref{fig:moneyplot}. Bin sizes reflect experimental resolution: \( \Delta E_e/E_e = 0.02 \oplus 0.15/\sqrt{E_e} \), \( \Delta\theta_e = 10\,\mathrm{mrad} \).

As shown in Fig.~\ref{fig:el}, despite the broad flux, the signal remains distinguishable: the cross section enhancement for heavy dark states boosts the signal at low \( E_e \), while moving away from the forward region suppresses background more effectively than signal. These features are explored semi-analytically in App.~\ref{app:eNscat}.

\section{Scattering against nucleons}\label{sec:Nscatt}

\begin{figure}
     \centering
     \includegraphics[width=1\linewidth]{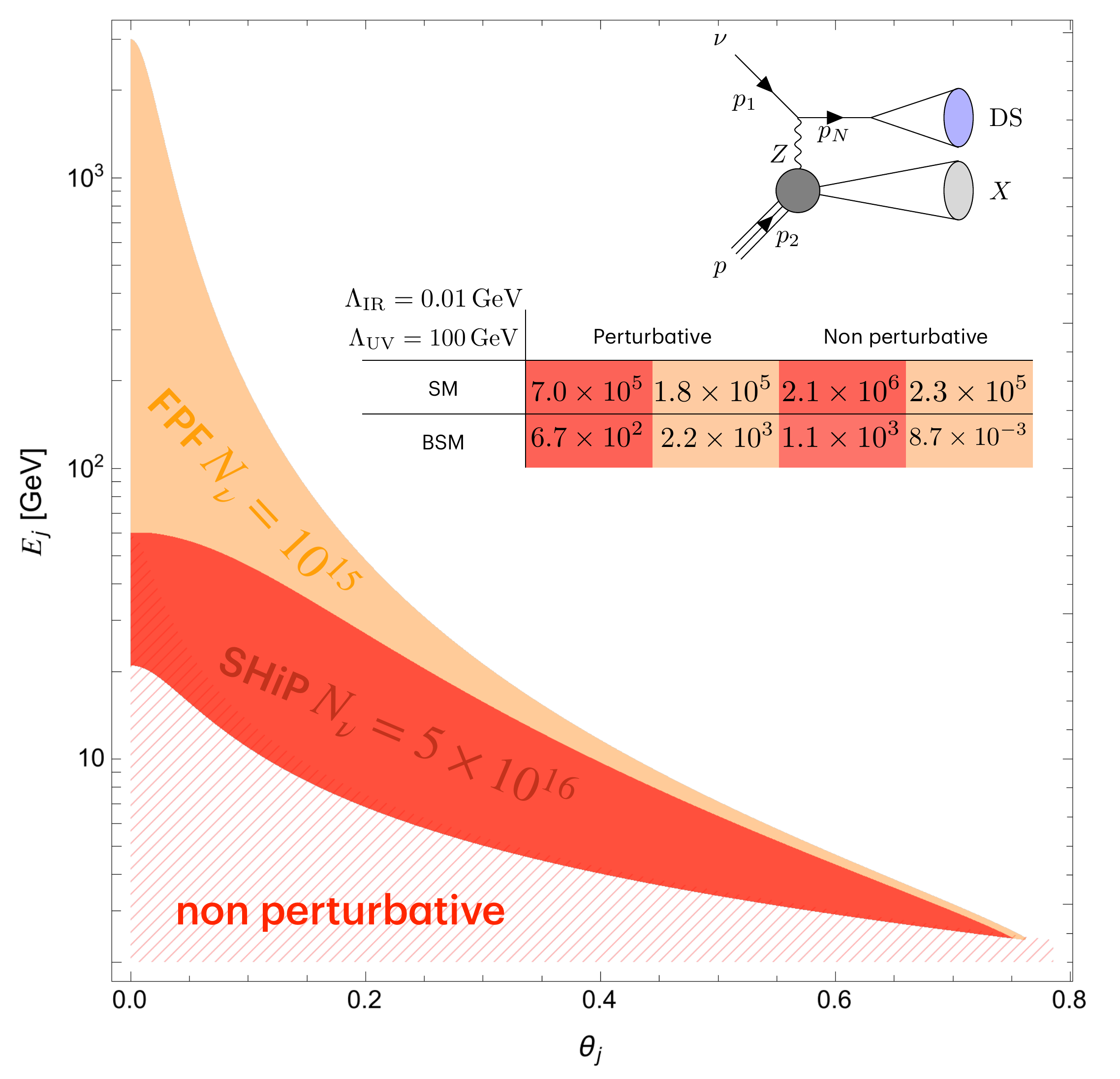}
     \caption{Allowed kinematical region for DIS at SHiP assuming an hypothetical flux with 10\% energy spread ({\bf red}) and at FASER-FPF ({\bf orange}) with the flux shown in \cref{fig:fluxes}. The {\bf hatched red} region shows the region of non-perturbative DIS events at SHiP, which is similar to the one at FPF. The lower and upper boundary of the perturbative region originate respectively from the requirements $Q^2>2m_p^2$ and $m^2_{\rm j}>2m_p^2$.  In the table, we show how background and signal events populate the perturbative and the nonperturbative region in the two experiments.}
     \label{fig:N}
 \end{figure}

In the nucleon scattering scenario, the signal arises from double deep inelastic scattering (2DIS) events in which both a visible and a dark-sector jet are produced in neutrino--nucleon interactions. Relative to the electron recoil case (see Sec.~\ref{sec:escatt}), the signal cross section is enhanced by a factor
\begin{equation}
\frac{s_p^2}{s_e^2} \sim 10^6 \left(\frac{E_\nu}{3~\mathrm{GeV}}\right)^2,
\end{equation}
reflecting the larger partonic center-of-mass energy available when scattering off nucleons.

The dominant background arises from SM neutrino DIS. In the forward limit, at fixed $E_\mathrm{j}$, $\theta_\mathrm{j}$ and for a given initial neutrino energy $E_\nu$, the exchanged momenta and jet masses for SM and BSM processes are approximately
\begin{align}
Q^2_{\rm SM} &\approx \frac{E_\nu E_{\rm j}^2 \theta_{\rm j}^2}{E_\nu - E_{\rm j}}, \ 
Q^2_{\rm BSM} \approx Q^2_{\rm SM} + \frac{m_N^2 E_{\rm j}}{E_\nu - E_{\rm j}}, \\
m^2_{{\rm j}, {\rm SM}} &\approx 2 m_p E_{\rm j}, \ 
m^2_{{\rm j}, {\rm BSM}} \approx m^2_{{\rm j}, {\rm SM}} - \frac{m_N^2 E_{\rm j}}{E_\nu - E_{\rm j}}\,.
\end{align}
These expressions imply \( Q^2_{\rm BSM} > Q^2_{\rm SM} \), ensuring that the signal remains in the perturbative regime even in the forward region where SM DIS becomes non-perturbative $(Q^2_{\rm SM}<2\,\rm{GeV}^2)$. This feature could potentially lead to a measurable difference in the jets substructure in the two scenarios that could help rejecting background.\footnote{We thank Jesse Thaler for enlightening discussions on this point.} 
While BSM jets are produced with lower invariant masses than their SM counterparts at fixed energy and angle, limited resolution in \( E_{\rm j} \), \( \theta_{\rm j} \), and \( m_j^2 \) constrains the discriminating power of this feature. We therefore focus on the inclusive enhancement of the neutral current cross section in our analysis.

We first constrain the signal using existing neutral current (NC) data from NuTeV. Adopting the flux and luminosity values in Table~\ref{tab:exp}, we conservatively require that signal events with visible jet energy \( E_{\rm jet} > 20\,\mathrm{GeV} \) — corresponding to $S = 1.7 \times 10^3 \left( \frac{100\,\mathrm{GeV}}{\Lambda_{\rm UV}} \right)^4$
— do not exceed the observed \( 4.57 \times 10^5 \) NC events~\cite{Zeller:2002he} (see Fig.~\ref{fig:moneyplot}).\footnote{The quoted luminosity includes acceptance and efficiency, extracted from Ref.~\cite{Zeller:2002he} via the DIS cross section above \( E_{\rm jet} > 20\,\mathrm{GeV} \). We checked that a refined binned-\( E_{\rm j} \) analysis improves the bound by only \(\sim 10\%\).} The jet energy cut ensures perturbative control over the SM background but is stricter than the NuTeV ``short track'' trigger. Approximately 30\% of signal events fail this requirement; we therefore relax this cut in projections for future experiments. In Fig.~\ref{fig:moneyplot}, we also show a constraint based on the quoted uncertainties in individual NuTeV energy bins. This constraints remains competitive with projections from upcoming experiments, motivating a reanalysis of the NuTeV data to correctly accounts for experimental efficiencies on the signal yield.

We next investigate the sensitivity of SHiP, using the beam parameters in Table~\ref{tab:exp}.\footnote{As shown in App.~\ref{app:nucleon}, our result is only mildly sensitive to the detailed neutrino flux, provided the mean energy is fixed at \( \langle E_\nu \rangle = 42\,\mathrm{GeV} \)~\cite{Pastore:2020dgg}.} We perform a two-bin analysis, separating kinematic regions where the SM background is perturbative from those where it is not. The red band in Fig.~\ref{fig:moneyplot} shows the reach from the purely perturbative bin (lower edge) and from the inclusive analysis including the non-perturbative region (upper edge).

We show an analogous projection for the FASER-FPF, where the higher \( \sqrt{s} \) enhances the signal relative to background and reduces the impact of non-perturbative regions.

\section{Conclusion}

Neutral current neutrino scattering provides a novel and powerful probe of composite sectors coupled to active neutrinos below the electroweak scale. We have identified scenarios where this approach outperforms conventional searches involving electroweak bosons or Higgs decays---particularly in models with low confinement scales, $\Lambda_{\rm IR} \gtrsim 10\,\mathrm{MeV}$, a continuum of strongly coupled states extending up to tens of GeV and an asymptotically free behavior at higher energy. 

Our analysis focused on inclusive enhancements to the neutral current cross section in UV-dominated regimes, where the dark-sector jet escapes the detector. Additional differential observables---such as hadronic energy deposition, angular distributions, and jet substructure---could further improve sensitivity, provided sufficient experimental resolution. 

In addition to beam neutrinos one could consider atmospheric neutrinos disintegrating in the earth. In this case the produced invisible jet will have the possibility of decaying back into SM particles inside the detector similarly to what has been considered in Refs.~\cite{Plestid:2020vqf,Plestid:2020ssy}. The jet decay will produce an extremely distinguishable event characterized by multiple charged tracks given the large number of single particle states expected from the hadronization of the invisible jet.

Many of the considerations related to neutrino disintegration discussed here are expected to carry over to IR-dominated portals. In this regime, dark-sector resonances are likely to decay within the detector, potentially producing distinctive emerging jet signatures at neutrino experiments~\cite{Schwaller:2015gea,Linthorne:2021oiz}. Of course in this scenarios complementary bounds from standard beam dump searches should also be considered~\cite{CHARM:1983ayi}. A detailed modeling of such scenarios, incorporating both resonant and continuum states, is a promising direction we aim to explore in future work~\cite{future}.

\section*{Acknowledgments}
We thank Adi Ashkenazi, Walter Marcello Bonivento, Lorenzo Di Pietro, Roni Harnik, Shirley Li, Maxim Pospelov, Andrea Tesi and Jesse Thaler for useful discussions on various aspects of this paper. The work of DR is supported in part by the European Union - Next Generation EU through the PRIN2022 Grant n. 202289JEW4. The work of DR and MC was performed in part at the Aspen Center for Physics, which is supported by National Science Foundation grant PHY-2210452. MC is supported in part by Perimeter Institute for Theoretical Physics. Research at Perimeter Institute is supported by the Government of Canada through the Department of Innovation, Science and Economic Development Canada and by the Province of Ontario through
the Ministry of Research, Innovation and Science.

\appendix 
\onecolumngrid
 \section{Two-point CFT correlators in momentum space}\label{eq:CFT}
We review the basic idea pioneered in Refs.\cite{Georgi:2007ek,Georgi:2007si}, which relates CFT correlation functions to scattering amplitudes. While conformal field theories do not admit a standard S-matrix due to the absence of asymptotic momentum eigenstates, it is still possible to define meaningful scattering amplitudes. These appear as residues of singularities in the Fourier transform of time-ordered correlation functions in Lorentzian signature\cite{Gillioz:2020mdd}.

For simplicity, let us consider a scalar operator $\mathcal{O}_S(x)$ of dimension $\Delta_S\geq1$ sourced by a source $\Phi(x)$
\begin{equation}
\Delta S=\int d^4x \Phi(x) \mathcal{O}_S(x)\,.
\end{equation}
The operator $\mathcal{O}_S(x)$ interpolates dark states $\vert N\rangle$ from the vacuum $\vert0\rangle$. The optical theorem relates the inclusive production rate of these states to the imaginary part of the two-point function: 
\begin{equation}
    \sum_N\int\Phi_{\rm DS}\vert\langle0\vert O_S\vert N\rangle\vert^2=2{\rm{Im}}\left[i\langle T[\mathcal{O}_S(p) \bar{\mathcal{O}}_S(-p)]\rangle\right]\,.
\end{equation}
This identity allows the two-point function to encode the relevant information for neutrino scattering into dark-sector states, as long as the process is inclusive and lies within the conformal regime ($\Lambda_{\rm IR} < \sqrt{s} < \Lambda_{\rm UV}$).

In Euclidean signature, the conformal two-point function in position space is fixed by symmetry:
\begin{equation}
\langle\mathcal{O}_S(x)\mathcal{O}_S(0)\rangle=\frac{c_{S}}{4\pi^2}\frac{1}{x^{2\Delta_S}}\ ,\label{eq:coordinate2point}    
\end{equation}
where the normalization is fixed to reproduce the massless scalar propagator for $\Delta_S=1$ when $c_{S}=1$, and the central charge $c_{S}>0$ is positive definite as required by unitarity. With this normalization, the central charge encodes the number of degrees of freedom interpolated by the operator $\mathcal{O}_S(x)$. Fourier transforming the 2-point function in Eq.~\eqref{eq:coordinate2point} we get 
\begin{equation}
\langle\mathcal{O}_S(p)\mathcal{O}_S(-p)\rangle=\int d^4x e^{ipx}\langle\mathcal{O}_S(x)\mathcal{O}_S(x)\rangle=\frac{c_{S}\Gamma(2-\Delta_S)}{2^{2\Delta_S-2}\Gamma(\Delta_S)}p^{2\Delta_S-4}\ .\label{eq:fourier}
\end{equation}
This result is valid for $1 \leq \Delta_S < 2$, while for $\Delta_S > 2$ the Fourier transform diverges and requires regularization. Nonetheless, the expression can be extended to general $\Delta_S$ by analytic continuation. A special treatment is needed when $\Delta_S = 2 + k$ with $k \in \mathbb{N}$, since $\Gamma(2 - \Delta_S)$ develops poles and the correlator becomes local in momentum space~\cite{Grinstein:2008qk,Bzowski:2013sza}.
This case is most easily treated in momentum space~(see for example Ref.~\cite{Bzowski:2013sza}). We regularize the divergence using dimensional regularization and renormalize the two-point function by introducing a local counterterm in the source $\Phi$. Shifting the spacetime dimension from $4$ to $4 - 2\epsilon$, the regulated two-point function consistent with dilatation Ward identity becomes:
\begin{equation}
    \langle\mathcal{O}_S(p)\mathcal{O}_S(-p)\rangle\vert_{\rm{reg}}=c(\epsilon)p^{2\Delta_S-4+2\epsilon}\,,
\end{equation}
where the normalization is a function of the regulator $\epsilon$ with  at most a simple pole because of locality of UV divergences and has a power expansion for $\epsilon\ll1$ defined as \begin{equation}
    c(\epsilon)\simeq \frac{\bar{c}}{\epsilon}+c'+\dots\,.
\end{equation} 
Expanding near $\epsilon \to 0$, we obtain
\begin{equation}
\langle\mathcal{O}_S(p)\mathcal{O}_S(-p)\rangle\vert_{\rm{reg}}\simeq p^{2\Delta-4}\left[\bar{c}\left(\frac{1}{\epsilon}+\log p^2\right)+c'\right]\,.
\label{eq:epsilonpole}
\end{equation}
The pole can be subtracted by adding a  local counter term in the source action
\begin{equation}
\Delta S_{\rm counter}(\Phi,\epsilon,\mu)=c_{\Phi}(\epsilon)\int d^{4-2\epsilon}x \mu^{2\epsilon} \Phi \Box^k\Phi\,.\label{eq:counter2}
\end{equation}
where the coefficient takes the form
\begin{equation}
c_\Phi(\epsilon)\simeq (-)^{k+1}\frac{\bar{c}}{2\epsilon}+c_\Phi'+\dots\,.
\end{equation}
After renormalization the 2-point function reads
\begin{equation}
\langle\mathcal{O}_S(p)\mathcal{O}_S(-p)\rangle\vert_{\rm{reg}}=p^{2\Delta-4}\left(\bar{c}\log \frac{p^2}{\mu^2}+c'-c_\Phi'\right)\ ,
\end{equation}
where the constant piece is scheme dependent and can be absorbed into the renormalization scale $\mu$. To match the position-space normalization in Eq.~\eqref{eq:coordinate2point}, we fix
\begin{equation}
    \bar c=\frac{c_S(-1)^{\Delta_S-1}}{\Gamma(\Delta_S)\Gamma(\Delta_S-1)4^{\Delta_S-1}}\,.
\end{equation}

After rotating to Minkowski space via the $i\epsilon$ prescription, the time-ordered momentum-space correlator becomes:
\begin{equation}
   \langle T\mathcal{O}_S(-p)\mathcal{O}_S(p)\rangle=\begin{cases}
       \frac{-ic_S\Gamma(2-\Delta_S)}{\Gamma(\Delta_S)4^{\Delta_S-1}}(-p^2-i\epsilon)^{\Delta_S-2}\,, \quad {\rm for } ~\Delta_S\neq 2+k\\
       \frac{-ic_S(-1)^{\Delta_S}}{\Gamma(\Delta_S)\Gamma(\Delta_S-1)4^{\Delta_S-1}}(-p^2-i\epsilon)^{\Delta_S-2}\log \frac{-p^2-i\epsilon}{\mu^2}\,,\quad {\rm otherwise.}
   \end{cases} \,. \label{eq:momspace2pt}
\end{equation}
Taking the imaginary part for use in the optical theorem yields:
\begin{equation}
  {\rm Im}(i \langle T\mathcal{O}_S(-p)\mathcal{O}_S(p)\rangle)=c_{\mathcal{O}_S}\frac{\pi(\Delta_S-1)}{\Gamma(\Delta_S)^24^{\Delta_S-1}}(p^2)^{\Delta_S-2}
  \label{eq:imm2pt}
\end{equation}
which is an analytic function for any $\Delta_S\geq 1$. 

The same procedure applies to Dirac fermion operators $\mathcal{O}_N$ of scaling dimension $\Delta_N$. The two-point function is obtained by shifting $\Delta_S \to \Delta_N + 1/2$ and multiplying by an additional factor of 2 to reproduce the free fermion normalization~\cite{Osborn:1993cr}. The resulting imaginary part is: 
\begin{equation}
    {\rm Im}(i\langle T \mathcal{O}_N(-p)\bar{\mathcal{O}}_N(p)\rangle)=\slashed{p}\frac{c_N2^{3-2\Delta_N}\pi}{\Gamma(\Delta_N-3/2)\Gamma(\Delta_N+1/2)}(p^2)^{\Delta_N-5/2}\,,
\end{equation}
which matches the result of \cref{eq:2point}.

\section{Meson and EW gauge bosons decays into invisible jets}\label{app:meson}
We derive the partial decay widths of light mesons \(\mathbf{m}\), electroweak gauge bosons, and the SM Higgs into dark sector jets induced by the interaction in ~\cref{eq:compositeN}.

The decay width of a charged meson, \(\mathbf{m}^\pm \to \ell^\pm + \mathrm{MET}\), differential in the invariant mass of the invisible jet, is given by
\begin{equation}
   \frac{d \Gamma(\mathbf{m}^\pm \to \ell^\pm + \mathrm{dark\,jet})}{d m_N^2} = \frac{y_\mu^2 A_N}{32\pi^2} \frac{|V_{qq'}|^2 f_\mathbf{m}^2 m_\mathbf{m}^3}{v^2 \Lambda_{\rm UV}^4} \left[ \frac{m_N^2}{\Lambda_{\rm UV}^2} \right]^{\Delta_N - 7/2} \left[ \frac{m_N^2}{m_\mathbf{m}^2} \left(1 - \frac{m_N^2}{m_\mathbf{m}^2} \right) + \frac{m_\ell^2}{m_\mathbf{m}^2} \left( 1 + 2\frac{m_N^2}{m_\mathbf{m}^2} - \frac{m_\ell^2}{m_\mathbf{m}^2} \right) \right] |\mathbf{p}_\ell|\,,
   \label{eq:mesondecayw}
\end{equation}
where \( V_{qq'} \) is the CKM matrix element corresponding to the quark flavor transition, and
\begin{equation}
   |\mathbf{p}_\ell| \equiv \sqrt{ \left(1 - \frac{m_N^2}{m_\mathbf{m}^2} \right)^2 + \frac{m_\ell^2}{m_\mathbf{m}^2} \left( \frac{m_\ell^2}{m_\mathbf{m}^2} - 2\frac{m_N^2}{m_\mathbf{m}^2} - 2 \right) }
\end{equation}
is the magnitude of the lepton’s three-momentum in the meson rest frame.

It is noteworthy that for \( m_\ell \ll m_N \lesssim m_{\mathbf{m}} \), as in the case of electrons, the decay is dominated by the first term in the square brackets of Eq.~\eqref{eq:mesondecayw}. This term becomes UV-dominated for scaling dimensions \( \Delta_N \geq \frac{5}{2} \), as first computed in Ref.~\cite{Chacko:2020zze}. Conversely, for heavier leptons such as the muon, the kinematics only permits the production of the lightest dark sector resonances. In this regime, the decay width is dominated by the second term in the brackets, which becomes UV-dominated for \( \Delta_N \geq \frac{7}{2} \).

\begin{figure}[t!]
    \centering
    \includegraphics[width=0.5\linewidth]{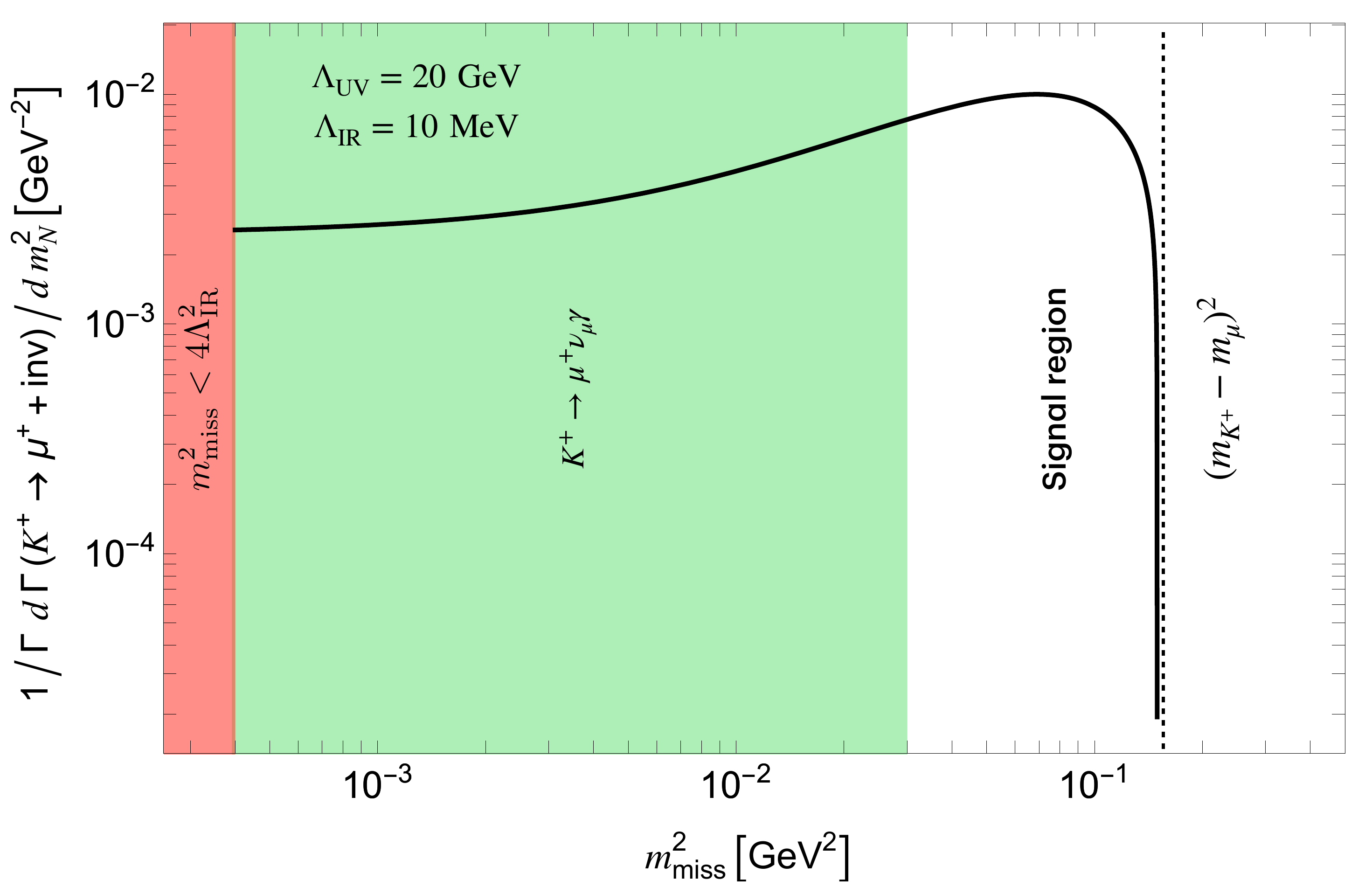}
    \caption{Normalized differential decay width \( \frac{1}{\Gamma} \frac{d\Gamma}{dm_N^2} \) for the process \( K^+ \to \mu^+ + \) invisible jet, for \( \Lambda_{\rm UV} = 20\,\mathrm{GeV} \) and \( \Lambda_{\rm IR} = 10\,\mathrm{MeV} \). The \textbf{red shaded region} indicates invariant masses too low to produce invisible jets. The \textbf{green shaded region} (\( m^2_{\rm miss} < 0.03\,\mathrm{GeV}^2 \)) is dominated by the SM background from \( K^+ \to \mu^+ \nu_\mu \gamma \) with soft photons (\( E_\gamma < 10\,\mathrm{MeV} \)). At higher invariant masses, SM backgrounds are suppressed~\cite{Cenci:2019nho}. Unlike narrow peaks from heavy neutral leptons, the signal exhibits a broad distribution only mildly peaked at high invisible jet invariant masses.}
    \label{fig:kplot}
\end{figure}

In practice, the latter regime is relevant for the results shown in Fig.~\ref{fig:moneyplot}. The constraints there are derived from current uncertainties in the SM branching ratio, although stronger bounds could be obtained by searching for the flat missing invariant mass distribution characteristic of invisible jets, as illustrated in Fig.~\ref{fig:kplot} for charged kaon decays. We encourage experimental collaborations to pursue such analyses. 

The decay widths of SM EW bosons and the Higgs can be written as
\begin{align}
   \frac{ d\Gamma(Z\to \nu+{\rm dark\, jet})}{dm_N^2}&=\frac{A_Ny_\mu^2}{48\pi^2}\frac{m_Z^3}{\Lambda_{\rm UV}^4}\left[\frac{m_N^2}{\Lambda_{\rm UV}^2}\right]^{\Delta_N-7/2}\left[1-\frac{m_N^2}{m_Z^2}\right]^2\left[2+\frac{m_N^2}{m_Z^2}\right]\,,\\
     \frac{    d\Gamma(W^\pm\to \ell^\pm+{\rm dark\, jet})}{dm_N^2}&=\frac{A_Ny_\mu^2}{24\pi^2}\frac{m_W^3}{\Lambda_{\rm UV}^4}\left[\frac{m_N^2}{\Lambda_{\rm UV}^2}\right]^{\Delta_N-7/2}\left[1-\frac{m_N^2}{m_W^2}\right]^2\left[2+\frac{m_N^2}{m_W^2}\right]\,,\\
     \frac{    d\Gamma(h\to \nu+{\rm dark\, jet})}{dm_N^2}&=\frac{A_Ny_\mu^2}{32\pi^2}\frac{m_h}{\Lambda_{\rm UV}^2}\left[\frac{m_N^2}{\Lambda_{\rm UV}^2}\right]^{\Delta_N-5/2}\left[1-\frac{m_N^2}{m_h^2}\right]^2\, ,
\end{align}
where the first two partial width agrees with Ref.~\cite{Chacko:2020zze} while the last one is new and as we see from Fig.~\ref{fig:moneyplot} leads to stronger constraints on $\Lambda_{\rm{UV}}$. 

Defining a variable $\zeta\equiv m_N^2/m_{Z,W,h}^2$, the differential cross sections above can be integrated as
\begin{align}
    \int_\delta^1 \zeta^{\Delta_N-7/2}(2+\zeta)(1-\zeta)^2d\zeta&=\frac{2-2 \delta ^{\Delta_N+\frac{1}{2}}}{2 \Delta_N+1}+\frac{6-6 \delta ^{\Delta_N-\frac{3}{2}}}{3-2 \Delta_N}-\frac{4-4 \delta ^{\Delta_N-\frac{5}{2}}}{5-2 \Delta_N}\,,\\
   \int_\delta^1 \zeta^{\Delta_N-5/2}(1-\zeta)^2d\zeta&= \frac{16 -2 \delta ^{\Delta_N-3/2} \left(4 {\Delta_N}^2 (\delta -1)^2-8 {\Delta_N} \delta  (\delta -1)+3 \delta  (\delta +2)-1\right)}{(2 {\Delta_N}-3) (2 {\Delta_N}-1) (2 {\Delta_N}+1) }\,.
\end{align}
where $\delta=\Lambda_{\rm IR}^2/m_{b}^2$ with $b=W,Z,h$ is the ratio between the square of the confinement scale and the square of the SM boson mass. Notice that the Higgs partial width depends differently on the invisible jet invariant mass compared to the EW bosons decays. This is because the production of a dark sector object does not need the insertion of  an off-shell neutrino propagation. As a consequence, the decay of EW (Higgs) bosons is UV dominated for $\Delta_N\geq 7/2$ ($\Delta_N\geq 5/2$). We stress that for the meson decays, the fact that $\mathcal{O}_N$ can excite a multiparticle state with invariant mass large compared to the mass of the single particle states $\Lambda_\mathrm{IR}$ makes it possible to avoid the large chirality suppression of order $\mathcal{O}(\Lambda_\mathrm{IR}/m_h)^2$. For completeness, we refer to \cref{tab:br} for the experimental values of the branching ratio used in this work.
\begin{table}[t!]

    \centering
    
    \begin{tabular}{c|c c }
          decay mode&\(\Gamma_{\rm tot}=\tau^{-1}\) [GeV] &  branching ratio      \\
        \hline
 \rowcolor{black!10}  \(\pi^+\to \mu^++{\rm inv} \) & \(2.524\times10^{-17}\) & \((99.98770\pm 0.00004)\) \% \\
\rowcolor{black!15} \(K^+\to \mu^++{\rm inv} \) &\(5.308\times10^{-17}\) & \((63.56\pm0.11)\) \%\\
 \rowcolor{blue!10}\(Z\to {\rm inv} \) & 2.4952 & \((20.000\pm0.055)\)\% \\
 \rowcolor{blue!25}  \(h\to {\rm inv} \)  &\(3\times10^{-3}\)  & \(<19\%\) \\

    \end{tabular}
    \caption{Decay modes, total decay widths of the decaying particles, and branching ratios considered in this work. The bounds derived for pions, kaons, and the $Z$ boson are obtained by requiring that the branching ratio to invisible jets remains below the experimental uncertainty on the measured branching ratio. For the Higgs boson, we impose that the contribution from dark decays to the invisible branching ratio does not exceed the current experimental upper limit. All the data are taken from Ref.~\cite{ParticleDataGroup:2022pth}.}
    \label{tab:br}
    \end{table}

\section{Details on the analysis}\label{app:eNscat}
In this section, we provide additional details on the derivation of the expected sensitivities for electron recoils at DUNE (Sec.~\ref{sec:electron}) and for nucleon recoils at NuTeV, SHiP, and FPF (Sec.~\ref{app:nucleon}).

\begin{figure}
    \centering
    \includegraphics[width=0.8\linewidth]{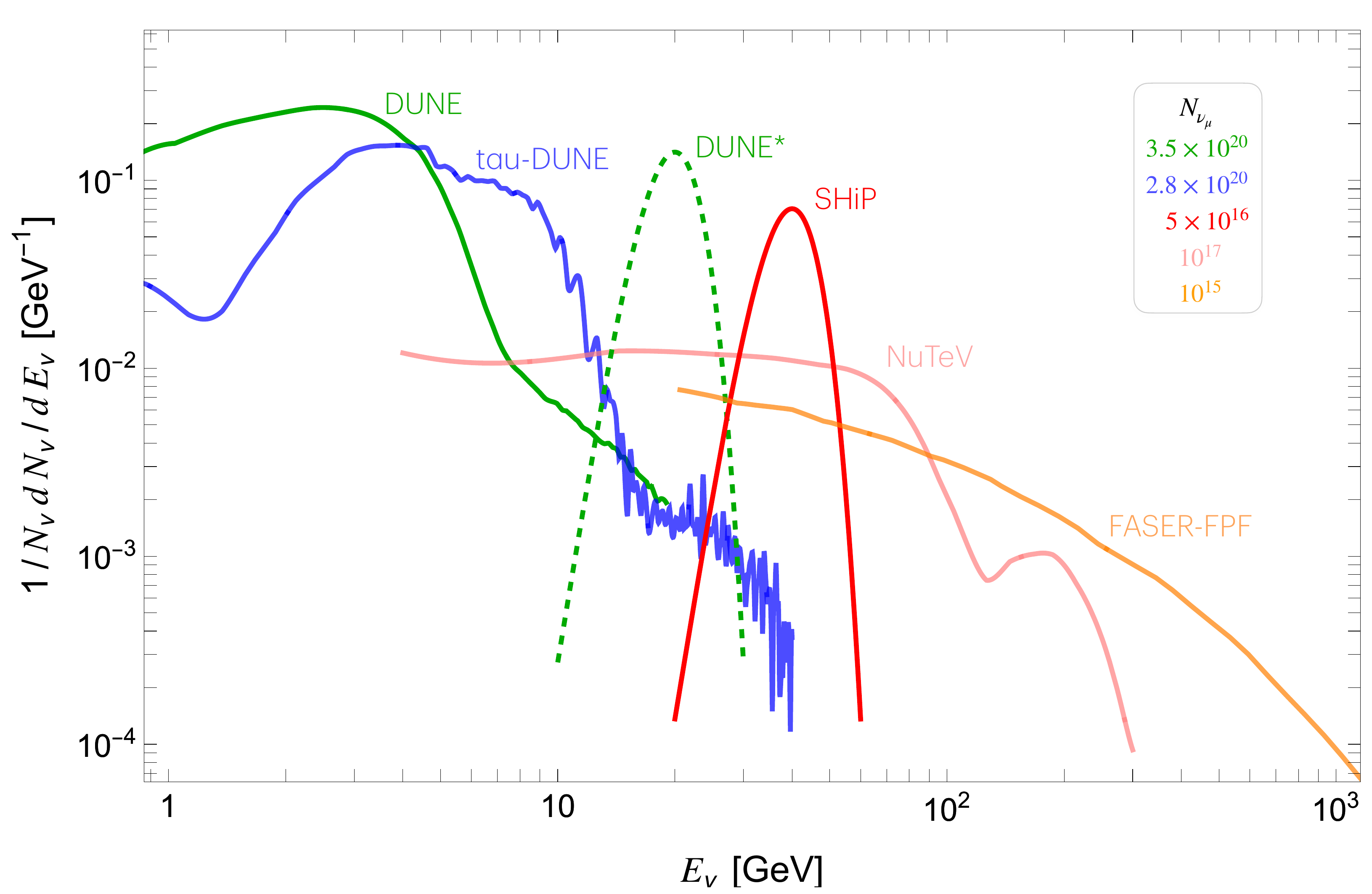}
    \caption{Summary plot of the normalized fluxes. {\bf Solid green:} realistic DUNE \(\nu_\mu\) flux~\cite{Ballett:2019bgd,DUNE:2020ypp}. {\bf Dashed green}: Gaussian flux (\(E_\nu = 20\) GeV, \(\sigma_E^2 = 8\) GeV\(^2\)) used in Fig.~4. {\bf Solid blue:} tau-optimized DUNE flux~\cite{Ballett:2019bgd}. {\bf Red:} SHiP Gaussian flux with (\(E_\nu = 40\) GeV, \(\sigma_E^2 = 32\) GeV\(^2\)). {\bf Solid pink: }NuTeV flux~\cite{NuTeV:2001whx}. {\bf Solid orange: }FASER-FPF flux~\cite{MammenAbraham:2024gun,Feng:2022inv}. FASER$\nu$ flux has the same shape of the one of FASER-FPF with a smaller luminosity of $N_\nu=10^{12}$.}
    \label{fig:fluxes}
\end{figure}

Given the central role of the neutrino energy flux in reconstructing both signal and background kinematics, we summarize in Fig.~\ref{fig:fluxes} the fluxes used in the sensitivity projections shown in Fig.~\ref{fig:moneyplot}. For DUNE, we consider both the baseline energy flux (solid green) and a tau-optimized flux~\cite{DUNE:2020ypp} (solid blue), the latter featuring a slightly enhanced high-energy tail. Given the outstanding expected resolution of DUNE in both energy and angle we also explore the potential of a more narrowly peaked neutrino beam (dashed green).

NuTeV's flux is broad and has a fat high energy tail~\cite{NuTeV:2001whx}. Similarly FASER$\nu$ and FASER-FPF which differ only for the expected number of collected neutrinos.  Since detailed information on the planned neutrino flux at SHiP is currently lacking, we model it as a Gaussian spectrum with a 20\% energy spread, as illustrated in Fig.~\ref{fig:fluxes}, and mean energy of 42 GeV~\cite{Pastore:2020dgg}. 

As we will discuss in Sec.~\ref{sec:electron}, the likelihood analysis for electron recoils is sensitive to both the mean energy and the width of the neutrino energy distribution. In contrast, as shown in Sec.~\ref{app:nucleon}, the nucleon recoil sensitivity is largely insensitive to the energy spread, due to the limited resolution on the hadronic jet energy, angle, and invariant mass, and is instead primarily determined by the mean energy and spectral shape of the neutrino flux.

\subsection{Electron recoils}\label{sec:electron}
For electron recoil events, the likelihood behavior can be understood analytically. We first recompute the signal rate in step by step. As in Eq.~\eqref{eq:diffe}, the matrix element squared averaged on the initial spins can be split into $|\bar M|^2=|\bar M_L|^2+|\bar M_R|^2$ where $M_L$ $(M_R)$ involves the left-handed (right-handed) electron current. These are 
\begin{equation}
|\bar M_L|^2=2\ell_e^2y_\mu^2A_{N}\frac{s_e(s_e-m_N^2)}{v^2\Lambda_{\rm UV}^4}\left(\frac{m_N}{\Lambda_{\rm UV}}\right)^{2\Delta_N-7}\,,\qquad |\bar M_R|^2=2r_e^2y_\mu^2A_{N}\frac{u_e(u_e-m_N^2)}{v^2\Lambda_{\rm UV}^4}\left(\frac{m_N}{\Lambda_{\rm UV}}\right)^{2\Delta_N-7}\,.        
\end{equation}
     
We now introduce the electron inelasticity variable $y_e$ where the momenta are labelled according to the diagram in Fig.~\ref{fig:el}. This reads
\begin{equation}
    y_e\equiv\frac{2p_2\cdot q}{s_e}=\frac{p_2\cdot (p_1-p_3)}{p_1\cdot p_2}=\frac{s_e+u_e-m_N^2}{s_e}\implies u_e=-s_e \left(1-y_e-\frac{m_N^2}{s_e}\right)\,.
    \label{eq:ydefelectron}
\end{equation}
We can write the fully differential cross-section as
\begin{align}
    d\sigma=\frac{2A_{N}y_\mu^2}{\Lambda_{\rm UV}^{4}v^2}\left[\frac{m_N^2}{\Lambda_{\rm UV}^2}\right]^{\Delta_N-7/2}\!\!\!\left[\ell_e^2(s_e-m_N^2)+r_e^2(1-y_e)(s_e-s_ey_e-m_N^2)\right] \frac{dm_N^2 dp_N^3dp_4^3}{(2\pi)^32E_N 2E_4}\delta^4(p_1+p_2-p_4-p_N)\,,
    \label{eq:elsignalxs}
\end{align}
where $l_e=1/2-\sin\theta_{\rm w}^2$, $r_e=\sin\theta_{\rm w}^2$ and $\sin\theta_{\rm w}^2\simeq0.22$. From now one we will focus on the left-handed electron current contribution for both signal and background but similar results can be obtained for the right-handed current.  

Resolving the DIS kinematics for the signal we can write the center of mass energy in the limit of high electron energies $E_e\gg m_e$ and small scattering angles $\theta_e\lesssim \sqrt{m_e/E_e}$
\begin{equation}
    s_e(E_e,\theta_e,m_N)= 2m_e E_\nu=\frac{m_e}{m_e-E_e(1-\cos\theta_e)}(m_N^2+2m_eE_e)\,.
\end{equation}
We can then fix the neutrino energy as a function of $m_N$, $E_e$ and $\theta_e$. We can then write the fully differential signal events distribution as a function of  electron energy, scattering angle, and invisible jet invariant mass:
\begin{equation}
    \frac{dS_L}{dm_N^2\,dE_e\,d\cos\theta_e} = \Phi(E_\nu)\sigma_{\rm{target}}^{-1}N_\nu \frac{A_N y_\mu^2 l_e^2}{4\pi^2 v^2 \Lambda_{\text{UV}}^4} \frac{E_e(s_e - m_N^2)}{2(m_e - E_e(1 - \cos\theta_e))} \,. \label{eq:ediffsign}
\end{equation}
Similarly, the background events distribution from neutrino-electron recoils is given in Ref.~\cite{Marciano:2003eq}  
\begin{equation}
    \frac{dB_L}{dE_e\,d\cos\theta_e} = \Phi(E_\nu)\sigma_{\rm{target}}^{-1} \frac{N_\nu}{\pi v^4} \frac{m_e E_e^2}{2(m_e - E_e(1 - \cos\theta_e))} l_e^2\, . \label{eq:ediffbkg}
\end{equation}

\begin{figure}[t!]
    \centering
    \includegraphics[width=0.5\linewidth]{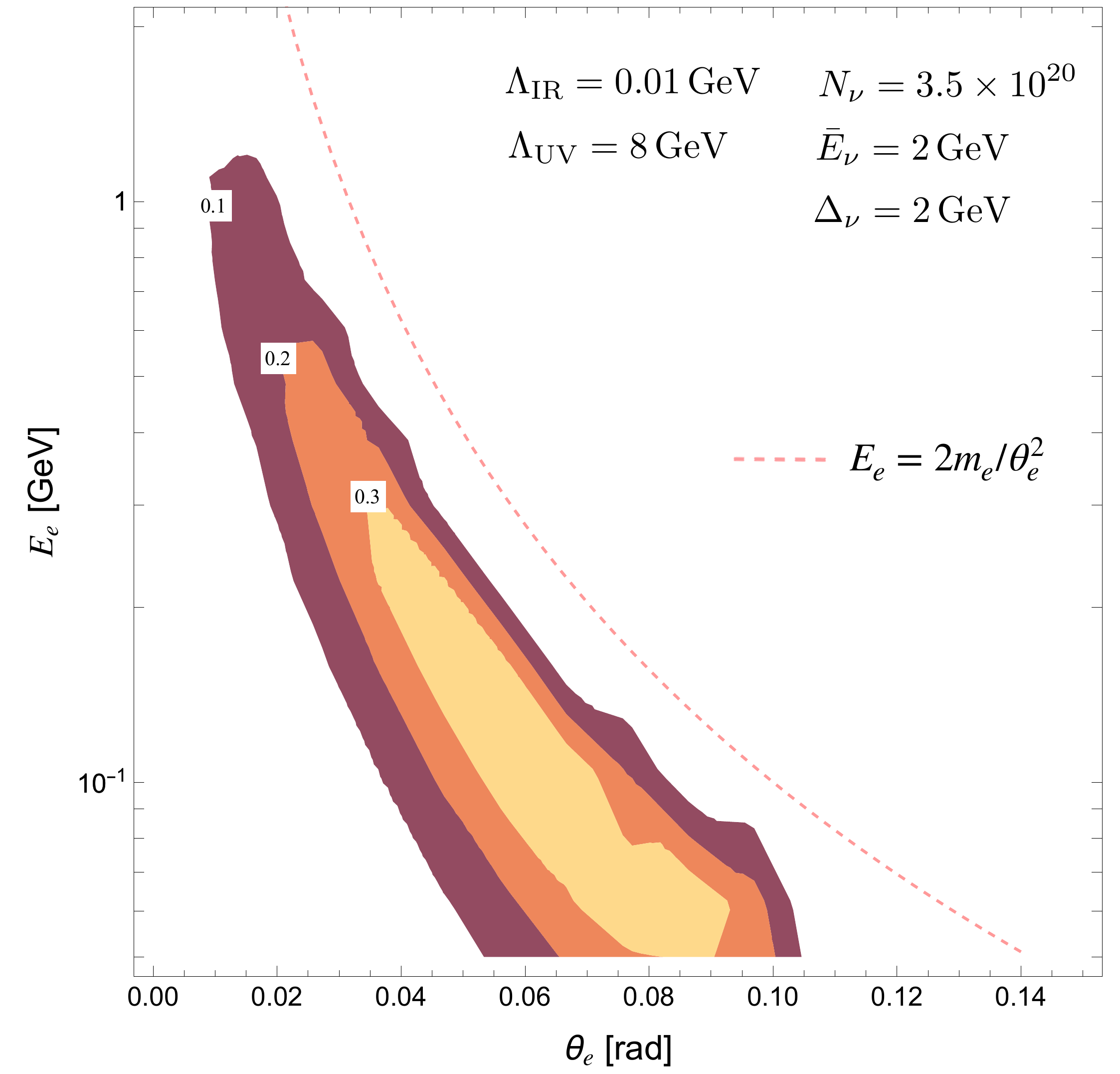}\hfill
    \includegraphics[width=0.5\linewidth]{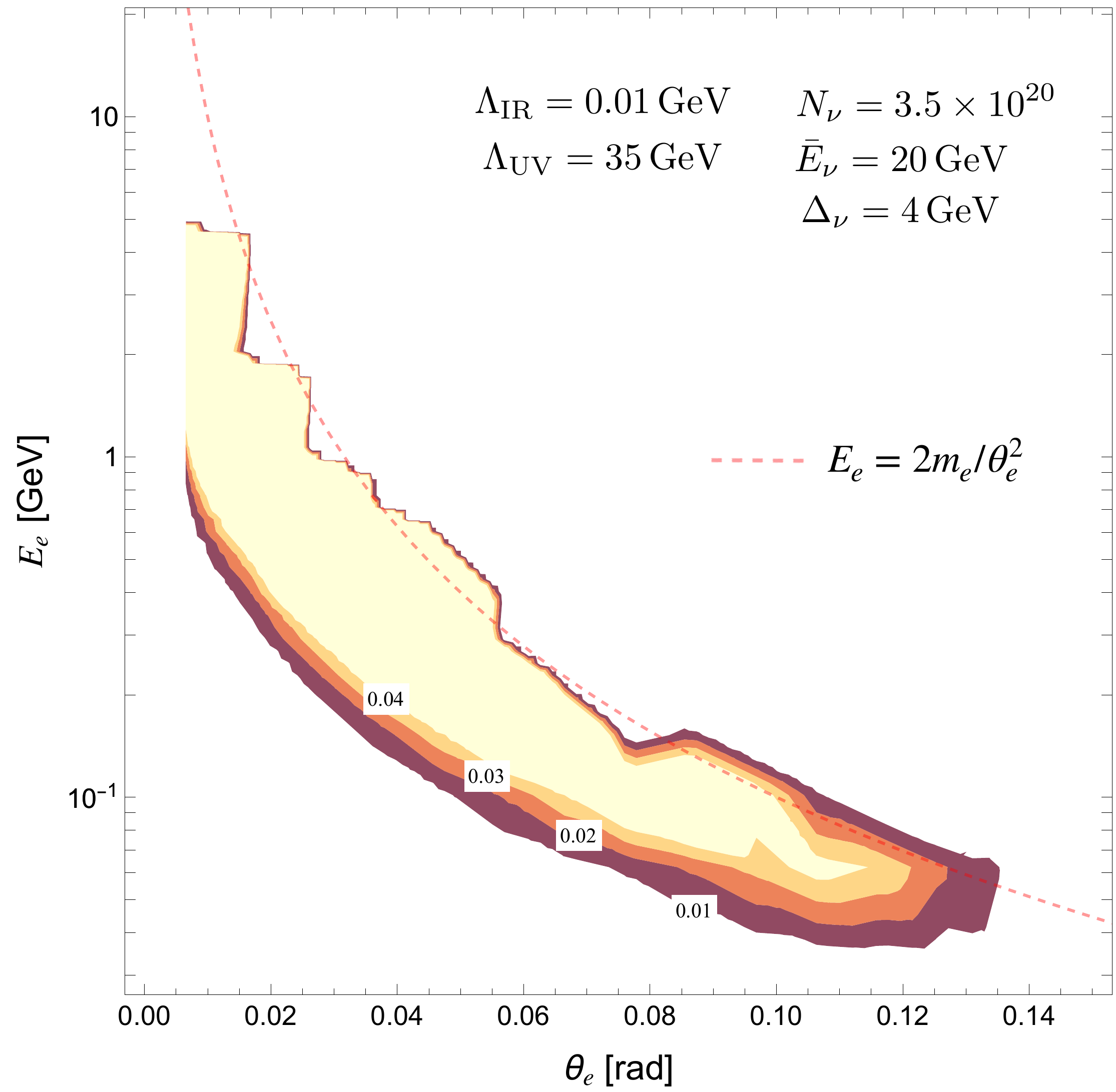}
     \caption{Test statistics $\lambda$ for neutrino scattering over electrons for different gaussian fluxes centered at $\bar E_\nu$ with width $\Delta_\nu$. In the gaussian approximation of the realistic flux scenario (left panel), the background significantly overlaps with the signal due to the broad energy spectrum of the incoming neutrinos. A key property to distinguish between signal and background arises from the interplay of kinematical and dynamical features: the signal decreases more slowly than the background at lower electron energies and larger scattering angles where the process is more inelastic and produces invisible jets with larger invariant masses. Although such events are kinematically suppressed, they are enhanced by the signal cross section $d\sigma_S/dm^2_N$. On the other hand, for the narrower and more energetic flux of DUNE$^*$ (right panel), kinematics enables a cleaner separation between signal and background, which works well also in the very forward region being guided by a tail hunt in $m_N^2$. 
}
    \label{fig:ellike}
\end{figure}

For a gaussian neutrino flux with mean energy $\bar{E}_\nu$ and variance $\Delta^2_\nu/2$ 
\begin{equation}
    \Phi(E_\nu)={\mathcal{N}}^{-1}{\exp\left(-\frac{(E_\nu-\bar{E}_\nu)^2}{\Delta_\nu^2}\right)}\,,\qquad {\mathcal{N}}=\sqrt\pi\Delta_\nu[1+\text{erf}(\bar{E}_\nu/\Delta_\nu)]/2\ ,
\end{equation}
it is possible to derive approximate analytical estimates of the likelihood integrating on $m_N^2$ and expanding in the forward region for $\theta_e\lesssim\sqrt{2m_e/E_e}$. In this limit, which is a good approximation for electron recoils as shown in Fig.~\ref{fig:ellike}, both signal and background are constant with respect to the scattering angle squared and can be integrated trivially. We are left with the differential distribution in the electron energy which can also be integrated analytically to obtain the number of events in the energy interval $[E_e(1 - R_e/2), E_e(1 + R_e/2)]$, where we defined $\Delta E_e / E_e \equiv R_e $ as the experimental resolution on the electron energy and analogously we define $R_\nu\equiv \Delta_\nu/E_\nu$ as the neutrino beam energy spread:
\begin{align}
    S_L(E_e)&=\frac{A_N y_\mu^2 l_e^2 N_\nu \sigma_{\rm{target}}^{-1} (2 m_e \bar E_\nu)^2}{16\pi^2\Luv^4 v^2 }F_S(E_e,\,R_e,\,R_\nu,\,\bar E_\nu)\\\notag&\approx 2\times 10^2 \left(\frac{y_\mu}{1}\right)^2\left(\frac{\bar E_\nu}{20\, {\rm GeV}}\right)^2\left(\frac{N_\nu}{10^{20}}\right)\left(\frac{\sigma_{\rm{target}}^{-1}}{0.02\,{\rm GeV}^2}\right)\left(\frac{\Luv}{10{\rm GeV}}\right)^4\left(\frac{F_S}{0.1}\right) \,,\\
    B_L(E_e)&=\frac{ l_e^2 N_\nu \sigma_{\rm{target}}^{-1}2m_e \bar E_\nu}{\pi^{3/2}v^4}F_B(E_e,\,R_e,\,R_\nu,\,\bar E_\nu)\approx 1.5\times 10^3\left(\frac{\bar E_\nu}{20\, {\rm GeV}}\right)\left(\frac{N_\nu}{10^{20}}\right)\left(\frac{\sigma_{\rm{target}}^{-1}}{0.02\,{\rm GeV}^2}\right)\left(\frac{F_B}{0.01}\right) \,,
\end{align}
with 
\begin{align}
F_S(E_e,\,R_e,\,R_\nu,\,\bar E_\nu)&\equiv\frac{1}{\text{erf}\left( 1/R_\nu \right) + 1} \Bigg\{ 
\frac{ \left( E_e^2 (R_e - 2)^2 - 2 \bar{E}_\nu^2( R_\nu^2+ 2  ) \right)
}{ \bar{E}_\nu^2 }\text{erf}\left( \frac{ -E_e R_e + 2 E_e - 2 \bar{E}_\nu }{ 2 R_\nu \bar{E}_\nu} \right) \notag \\[1ex]
&- \frac{ \left( E_e^2 (R_e + 2)^2 - 2 \bar{E}_\nu^2( R_\nu^2 + 2  ) \right)
}{ \bar{E}_\nu^2 } \text{erf}\left( \frac{ E_e (R_e + 2) - 2 \bar{E}_\nu }{ 2 R_\nu \bar{E}_\nu} \right)+ \frac{ 8 E_e^2 R_e }{ \bar{E}_\nu^2 } \notag \\[1ex]
&+ \frac{ 
2 R_\nu\left( 2 \bar{E}_\nu -  E_e (R_e - 2) \right)
}{ \sqrt{\pi} \bar{E}_\nu }\exp\left[ -\frac{(E_e (R_e - 2) + 2 \bar{E}_\nu)^2}{4 R_\nu^2\bar{E}_\nu^2} \right] \notag \\[1ex]
&- \frac{ 2 R_\nu (E_e (R_e + 2) + 2 \bar{E}_\nu)
}{ \sqrt{\pi} \bar{E}_\nu } \exp\left[ -\frac{(E_e (R_e + 2) - 2 \bar{E}_\nu)^2}{4 R_\nu^2\bar E_\nu^2} \right]\Bigg\}\,,\label{eq:fs}\\
    F_B(E_e,\,R_e,\,R_\nu,\,\bar E_\nu)&\equiv\frac{\sqrt{\pi} \left( \text{erf}\left( \frac{E_e (R_e - 2) + 2 \bar{E}_\nu}{2 R_\nu \bar{E}_\nu} \right) + \text{erf}\left( \frac{E_e (R_e + 2) - 2 \bar{E}_\nu}{2 R\nu\bar{E}_\nu} \right) \right)}{ \text{erf}\left( 1/R_\nu \right) + 1 } \notag\\[1ex]&+ \frac{{R_\nu}  \left( \exp\left[ -\frac{(E_e (R_e - 2) + 2 \bar{E}_\nu)^2}{4 R_\nu^2\bar{E}_\nu^2} \right] - \exp\left[ -\frac{(E_e (R_e + 2) - 2 \bar{E}_\nu)^2}{4 R_\nu^2\bar{E}_\nu^2} \right] \right)}{ \left[\text{erf}\left(1/R_\nu \right) + 1 \right]} \label{eq:fb}
\end{align}
dimensionless combinations of gaussian and error functions. Notice that the electron mass \( m_e \) enters the center-of-mass energy as \( s_e \sim 2 m_e \bar{E}_\nu \), leading to a suppression of the signal. This significantly limits the number of BSM-induced events and ultimately constrains the experimental sensitivity, even in high-luminosity setups.

In the following, we fix \(\sigma_{\rm target}^{-1} = 0.02\,\mathrm{GeV}^2\), \(y_\mu = 1\), and \(R_e = 0.05\). Eqs.~\eqref{eq:fs} and \eqref{eq:fb} exhibit interesting limiting behaviors that provide physical insight into how the invisible jet signature can be distinguished from Standard Model background. 

For instance, in the limit of a narrow neutrino flux \((R_\nu \ll 1)\), and expanding to first order in \(R_e \ll 1\), we obtain:
\begin{align}
    F_S &\approx \frac{8 E_e^2 R_e}{\bar{E}_\nu^2} \, \Theta_H(\bar{E}_\nu - E_e), \quad
    F_B \approx 2 \sqrt{\pi} \quad \text{for } E_e = \bar{E}_\nu,
\end{align}
where \(\Theta_H(x)\) is the Heaviside step function. In this regime, the initial-state kinematics can be accurately reconstructed, enabling a precise determination of the missing invariant mass \(m_N\), which becomes a well-defined observable due to the sharply known neutrino energy. This allows a clean separation between inelastic signal events \((m_N > 0)\) and elastic background \((m_N = 0,\, E_e \sim \bar{E}_\nu)\), where the final-state electron carries away most of the neutrino energy. 

In contrast, signal events typically transfer a significant fraction of the center-of-mass energy \(s_e\) to the invisible jet mass, leading to a less energetic outgoing electron compared to the background. As a result, a monochromatic neutrino beam yields a background-free signal except in the narrow region where \(E_e \approx \bar{E}_\nu\), where the elastic background is concentrated. In this regime, the likelihood analysis effectively becomes a search in the high-\(m_N\) tail:
\begin{equation}
    \lambda \approx S = \sum_i S_i \approx 2 \times 10^4 \left(\frac{N_\nu}{10^{20}}\right)\left(\frac{\Lambda_{\rm UV}}{\rm GeV}\right)^{-4} \sum_i \left(\frac{E_{e,i}}{\rm GeV}\right)^2,
\end{equation}
which is dominated by the bin with maximal electron energy \(E_{e,i} \sim \bar{E}_\nu\). Focusing on this high-energy bin, we estimate the sensitivity as:
\begin{equation}
    \Lambda_{\rm UV} \approx 12\,{\rm GeV} \left(\frac{N_\nu}{10^{20}}\right)^{1/4} \left(\frac{\bar{E}_\nu}{\rm GeV}\right)^{1/2}.
    \label{eq:ereachp}
\end{equation}

In the opposite regime of a very broad neutrino flux \((R_\nu \gg 1)\), the initial-state kinematics are less constrained, and there is no background-free region due to significant overlap between signal and background. However, a broader flux grants access to higher energies, substantially enhancing the signal yield. In this limit, we approximate:
\begin{align}
    F_S \approx \frac{8 E_e^2 R_e}{\bar{E}_\nu^2}, \quad
    F_B \approx \frac{2 E_e^2 R_e}{\bar{E}_\nu^2 R_\nu}.
\end{align}
In the high-statistics limit, the likelihood becomes a sum over signal-squared over background in each bin:
\begin{equation}
    \lambda \approx \sum_i \frac{S_i^2}{B_i} \approx 5 \times 10^5 \left(\frac{N_\nu}{10^{20}}\right)\left(\frac{R_\nu}{1}\right)\left(\frac{\bar{E}_\nu}{\rm GeV}\right)\left(\frac{\Lambda_{\rm UV}}{\rm GeV}\right)^{-8} \sum_i \left(\frac{E_{e,i}}{\rm GeV}\right)^2.
\end{equation}
Considering only the bin with maximum electron energy \(E_{e,\max} \approx R_\nu \bar{E}_\nu\), we estimate:
\begin{equation}
    \Lambda_{\rm UV} \approx 12\,{\rm GeV} \left(\frac{N_\nu}{10^{20}}\right)^{1/8} \left(\frac{R_\nu\bar{E}_\nu}{\rm 10\,GeV}\right)^{3/8}\,,
    \label{eq:ereachw}
\end{equation}
which depends only on $\Delta_\nu$.

There is also an intermediate regime where the flux width is comparable to its mean energy \((R_\nu \sim 1)\), resembling the realistic DUNE flux with \(\bar{E}_\nu \sim 3\,\mathrm{GeV}\). In this case, approximate forms for \(F_S\) and \(F_B\) are:
\begin{align}
    F_S &\approx \frac{4 E_e^2 R_e}{\bar{E}_\nu^2} \, \mathrm{erfc}\left(\frac{E_e}{\bar{E}_\nu} - 1\right), \quad
    F_B \approx \frac{E_e^2 R_e}{\bar{E}_\nu^2} \, \exp\left(-\frac{(E_e - \bar{E}_\nu)^2}{\bar{E}_\nu^2}\right).
\end{align}
Here, the main difference between signal and background arises from the complementary error function vs.\ exponential behavior, with the background falling off more rapidly as \(E_e\) decreases. In the high-statistics limit, the likelihood becomes:
\begin{equation}
    \lambda \approx \sum_i \frac{S_i^2}{B_i} \approx 3 \times 10^5 \left(\frac{N_\nu}{10^{20}}\right)\left(\frac{\bar{E}_\nu}{\rm GeV}\right)\left(\frac{\Lambda_{\rm UV}}{\rm GeV}\right)^{-8} \sum_i \left(\frac{E_{e,i}}{\rm GeV}\right)^2 \mathrm{erfc}^2\left(\frac{E_{e,i}}{\bar{E}_\nu}\right) \exp\left(\frac{(E_{e,i} - \bar{E}_\nu)^2}{\bar{E}_\nu^2}\right),
\end{equation}
which is maximized for \(E_{e,i} \lesssim \bar{E}_\nu\). Substituting \(E_{e,i} \sim \bar{E}_\nu\), the sensitivity becomes:
\begin{equation}
    \Lambda_{\rm UV} \approx 5\,{\rm GeV} \left(\frac{N_\nu}{10^{20}}\right)^{1/8} \left(\frac{\bar{E}_\nu}{\rm GeV}\right)^{3/8}.
    \label{eq:ereachr}
\end{equation}
This is parametrically the least favorable scenario among Eqs.~\eqref{eq:ereachp}, \eqref{eq:ereachw}, and \eqref{eq:ereachr}, as it neither benefits from a clean tail-based search in \(m_N\) nor from the signal enhancement due to high energies available in broad fluxes.

By comparing Eqs.~\eqref{eq:ereachp}, \eqref{eq:ereachw}, and \eqref{eq:ereachr}, we observe that, due to the steeper dependence on \(\bar{E}_\nu\) at fixed $R_\nu$, a tail hunt using a monochromatic neutrino flux \((\Lambda_{\rm UV} \propto \bar{E}_\nu^{1/2})\) with energies above a few GeV is more powerful than searches for distortions in the energy spectrum from broader fluxes \((\Lambda_{\rm UV} \propto \bar{E}_\nu^{3/8})\). Interestingly, Eqs.~\eqref{eq:ereachp} and \eqref{eq:ereachw} yield numerically similar sensitivities \(\Lambda_{\rm UV} \sim 10~\mathrm{GeV}\), roughly independent of the precise flux shape. However, the monochromatic-tail analysis benefits more from increased luminosity, with \(\Lambda_{\rm UV} \propto N_\nu^{1/4}\), compared to \(\Lambda_{\rm UV} \propto N_\nu^{1/8}\) in the broad-flux case.

\begin{figure}[t!]
    \centering
    \includegraphics[width=0.5\linewidth]{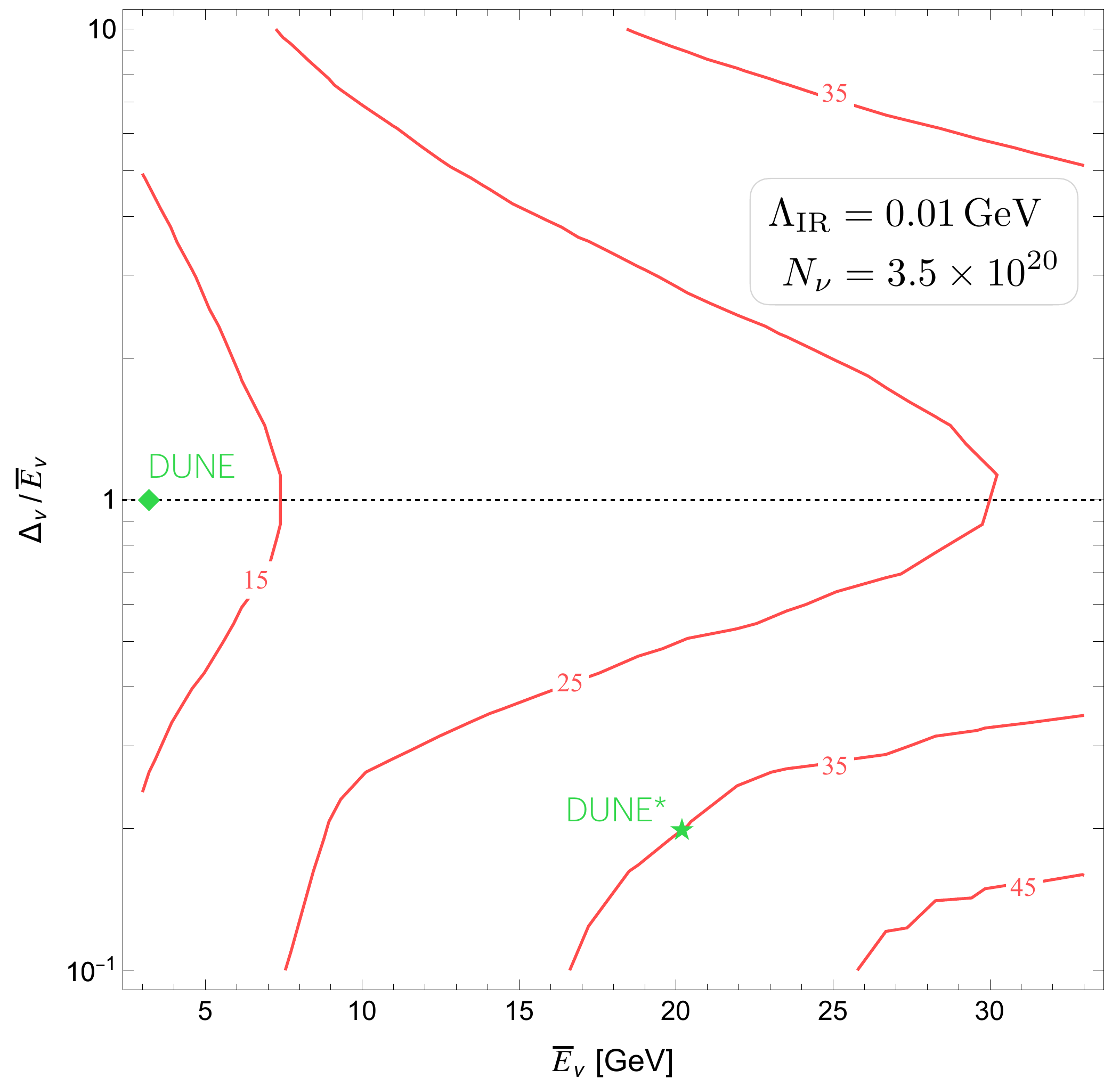}
    \caption{Expected sensitivities from scattering over electrons for an incoming gaussian flux centered at energy $\bar E_\nu$ with width $\Delta_\nu$. The numerical coincidence observed in \eqref{eq:ereachp}, \eqref{eq:ereachw}, and \eqref{eq:ereachr} leads to comparable values of $\Lambda_{\rm UV}$ when varying the flux shape at fixed $\bar E_\nu$. Furthermore, for big enough $\bar E_\nu$, a peaked flux, combined with a search in the missing invariant mass tail, provides the best sensitivity. Notice that the tail hunt tends to saturate as a function of \(\Delta_\nu / \bar{E}_\nu\) once the analysis becomes almost background-free. The green square points to a realistic DUNE-like experiment with a gaussian flux of mean energy $\bar E_\nu=3$ GeV and a width of $\Delta_\nu=\bar E_\nu$. The green star shows the DUNE$^*$ experiment, an ideal narrow monochromatic flux with  $\bar E_\nu=20$ GeV and  $\Delta_\nu=4$ GeV.}
    \label{fig:envsspread}
\end{figure}

The interplay between the mean energy and energy spread of the neutrino flux is illustrated in Fig.~\ref{fig:envsspread}. At fixed relative width \(R_\nu = \Delta_\nu / \bar{E}_\nu\), increasing \(\bar{E}_\nu\) enhances the signal rate, which grows faster than the background, thus improving sensitivity. Conversely, narrower fluxes improve the kinematic separation between signal and background, enabling a more precise distinction between elastic and inelastic events. Fig.~\ref{fig:ellike} shows two examples of the test statistic \(\lambda\): a DUNE-like scenario (left panel) and a high-energy monochromatic beam scenario (DUNE$^*$, right panel).


\subsection{Scattering against nucleons}\label{app:nucleon}

This section covers the signal and background processes for neutrino scattering off nucleons, focusing on protons. At momentum transfer \(Q^2 \gtrsim 2 m_p^2\), neutrino scattering involves free quarks inside nucleons, and QCD is perturbative, allowing for a partonic description. At lower \(Q^2\), the strong coupling increases, making partons no longer free, and non-perturbative effects dominate. To exclude events close to QCD the resonance region, we further require that the invariant mass of the hadronic jet $m^2_{\rm j}>2m_p^2$. For simplicity, we focus on proton targets; neutron results follow from isospin symmetry.

The invisible jet production cross-section is derived by replacing the electron in the previous section with a quark \(q\) that carries a fraction \(x\) of the proton’s momentum. The neutral current charges are modified to \(\ell_q\) and \(r_q\), where \(\ell_q = I_{3,q} - Q \sin^2\theta_W\) and \(r_q = -Q \sin^2\theta_W\) represent the left- and right-handed neutral current couplings, and \(I_{3,q}\) is the quark’s weak isospin.

\begin{figure}[t!]
    \centering
    \includegraphics[width=0.5\linewidth]{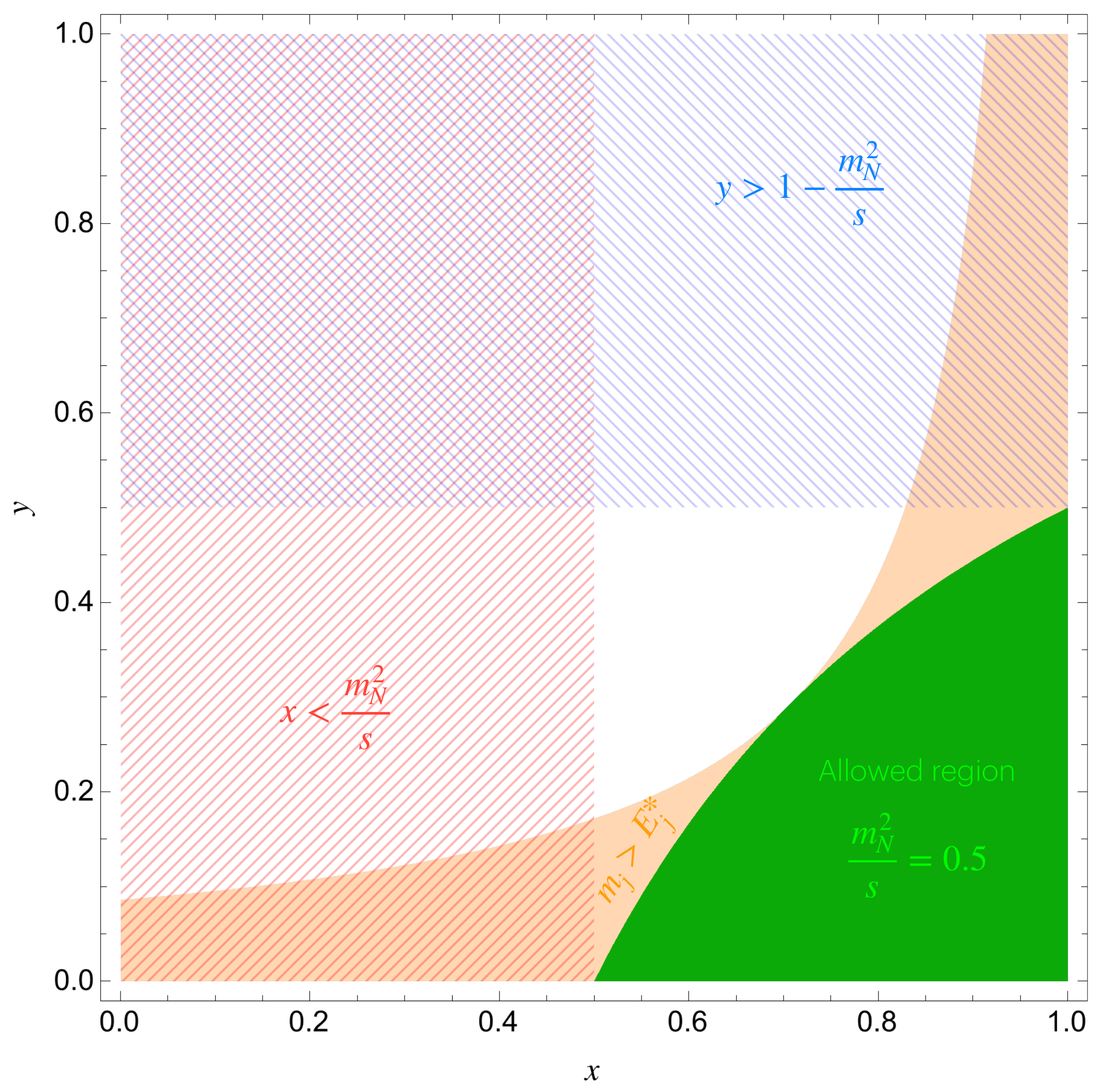}
    \caption{Kinematics in the (\(x,\,y\)) plane for \(m_N^2/s = 0.5\). The \textbf{hatched red} region is excluded because the initial state quark does not carry sufficient momentum to produce an invisible jet with mass \(m_N^2\). The \textbf{hatched blue} region corresponds to configurations where the hadronic jet energy is incompatible with the emission of an invisible jet of mass \(m_N^2\), and is therefore excluded. The \textbf{orange shaded} area satisfies the first condition in \eqref{eq:kinreg}, while the \textbf{green shaded} region fulfills the second, more stringent, condition, which ultimately determines the allowed kinematic configurations. A non negligible mass of the proton slightly modifies the shape of the allowed region close to $y=0$.}
    \label{fig:kinregxy}
\end{figure}

The center-of-mass energy of the partonic subprocess is \(\hat{s} = x s = 2 x m_p E_\nu\), and the subprocess is weighted by the quark PDFs \(f_q(Q^2, x)\). We express \(x\) and inelasticity \(y\) as:
\begin{equation}
x = \frac{Q^2}{2p_2 \cdot q}, \quad y = \frac{2p_2 \cdot q}{s} \implies Q^2 = x y s \ .   
\end{equation}
The fully differential cross-section for the partonic subprocess \(q + \nu \to q + \text{DS}\) is very similar to what was presented in Eq.~\eqref{eq:elsignalxs} for electrons:
\[
d\sigma = \frac{2A_N y_\mu^2}{\Lambda_{\rm UV}^4 v^2} \left(\frac{m_N^2}{\Lambda_{\rm UV}^2}\right)^{\Delta_N - 7/2} \sum_{q, \bar{q}} f_q(Q^2, x) \left[\ell_q^2(\hat{s} - m_N^2) + r_q^2(1 - y)(\hat{s} - \hat{s}y - m_N^2)\right] \frac{dm_N^2 dp_N^3 d\hat{p}_4^3}{(2\pi)^3 2 E_N 2 \hat{E}_4} \delta^4(p_1 + \hat{p}_2 - p_N - \hat{p}_4)\,.
\]
Integrating over the final state phase space at fixed $m_N^2$ the phase space simplifies to the following:
\[
\frac{dm_N^2}{16\pi^2} \frac{\hat{E}_4 d\hat{E}_4 d\cos\theta}{E_N} \delta(\hat{s} - E_N - \hat{E}_4)\,,
\]
where in the center-of-mass we have
\[
\hat{E}_4 = \frac{\hat{s} - m_N^2}{2\sqrt{\hat{s}}}, \quad E_N = \frac{\hat{s} + m_N^2}{2\sqrt{\hat{s}}}\,.\label{eq:kinreg}
\]
Solving the remaining Dirac-delta leads to the Jacobian $j= \frac{E_N}{\hat{E}_4}\frac{\hat{s} - m_N^2}{2\hat{s}}$.   

Defining \(\hat{t} = -Q^2 = (p_2 - p_4)^2\) and using the relation \(d\hat{t} = d\cos\theta \frac{\hat{s} - m_N^2}{2}\), we find $d x d\hat{t} = x s dx dy$. All in all, the differential cross-section as a function of \(x, y, m_N^2\) is
\[
\frac{d\sigma_S}{dx dy dm_N^2} = \frac{A_N y_\mu^2}{8\pi^2 \Lambda_{\rm UV}^4 v^2} \left(\frac{m_N^2}{\Lambda_{\rm UV}^2}\right)^{\Delta_N - 7/2} \sum_{q, \bar{q}} f_q(Q^2, x) \left[\ell_q^2(xs - m_N^2) + r_q^2(1 - y)(xs - xsy - m_N^2)\right]\,.
\]
The SM DIS differential cross-section is
\begin{equation}\label{eq:calcdisbkg}
  \frac{d\sigma_B^{\rm DIS}}{dx dy} = \frac{s}{2\pi v^4} \sum_{q, \bar{q}} x f_q(Q^2, x) \left[\ell_q^2 + r_q^2(1 - y)^2\right]  \,.
\end{equation}
The allowed region in the \(x, y\) plane for BSM events is smaller than for SM DIS. This is due to kinematic constraints, which arise from the conditions:
\begin{equation}
E_j^* \geq m_j, \quad \cos^2\theta_j^* \leq 1    \,,
\end{equation}
which lead respectively to the following constraints in the \(x, y\) plane:
\[
1 - \sqrt{y(1 - x)} \geq \sqrt{\zeta}, \quad x(1 - y) \geq \zeta\,,
\]
where \(\zeta = m_N^2 / s\). For instance the production of an invisible jet with mass \(m_N^2\) requires $x \geq \frac{m_N^2}{s}, \quad y \leq 1 - \frac{m_N^2}{s}$.

These constraints limit further the allowed region for invisible jet production, as illustrated in Fig.~\ref{fig:kinregxy} by the difference between the orange region satisfying Eq.~\eqref{eq:kinreg} and the green region which is the allowed one for $m_N^2=0.5\, s$. Of course the difference between the two region is maximized for heavier $m_N$ which is also where the signal cross section is maximized.

In scenarios with low-energy neutrino beams, such as at SHiP, it is crucial to properly account for the nonperturbative component of SM DIS, which provides the dominant contribution in the soft jet emission regime. We deal with this problem performing a change of variables from the natural DIS kinematic variables \((E_\nu,\, x,\, y,\, m_N^2)\)  to the laboratory-frame variables \((E_\nu,\, E_j,\, \cos\theta_j,\, m_N^2)\), via the following set of equations
\begin{align}
  m^2_j= E_j^2-\textbf{p}_j^2&=(1-x) y \left(2 E_\nu m_p+m_p^2\right)+m_p^2\,,\\y&=\frac{2 m_p \left(E_\nu-\sqrt{m_N^2+\textbf{p}_N^2}\right)}{2 E_\nu m_p+m_p^2}\,,\\m_N^2+\textbf{p}_N^2&=(-E_j+E_\nu+m_p)^2\,,\\\left(1-\cos\theta_j^2\right) \textbf{p}_j^2&=\left(1-\cos\theta_N^2\right) \textbf{p}_N^2\,,\\ E_\nu &=\cos\theta_N \vert \textbf{p}_N\vert+\cos\theta_j \vert\textbf{p}_j\vert\,.
\end{align}
Given the uncertainty on the energy of the incoming neutrino due to broad energy spectrum, the exchanged momentum transfer $Q^2$ cannot be directly reconstructed. To provide a reliable estimate of the sensitivity at SHiP, we discretize the hadronic jet phase space into bins centered at $(E_j, \theta_j)$, with bin widths determined by the detector resolution. We assume a relative energy resolution of $\Delta E_j / E_j = 30\%$ and an angular resolution of $\Delta\theta_j = 100$ mrad.  For each bin, we compute the momentum transfer using the maximum energy available in the neutrino flux $E_{\nu,{\rm max}}$ as
\begin{equation}
    Q^2(E_j,\theta_j,E_{\nu,{\rm max}}) = 2s_p(E_{\nu,{\rm max}})x(E_j,\theta_j,E_{\nu,{\rm max}})y(E_j,\theta_j,E_{\nu,{\rm max}})
\end{equation} 
and classify it as part of the nonperturbative regime if $0<Q^2(E_j,\theta_j,E_{\nu,{\rm max}})<2 m_p^2$. By summing all events in each region, we obtain the BSM and SM number of events in the perturbative region $N_{\rm BSM, pert},\,N_{\rm SM, pert}$, as well as the number of signal events in the nonperturbative region $N_{\rm BSM, nonpert}$, and the calculable contribution of SM events in the non perturbative regime $N_{\rm SM, nonpert}$. 

We also estimate the number of SM events in the region $Q^2(E_j,\theta_j,E) < 2\,\mathrm{GeV}^2$ where the the DIS calculation of Eq.~\eqref{eq:calcdisbkg} is no longer reliable. Starting from the inclusive CC cross-section as a function of $E_\nu$ from Ref.~\cite{Formaggio:2012cpf}, we obtain the corresponding NC cross-section by a rescaling with the ratio $R = \sigma_{\rm NC} / \sigma_{\rm CC}$ provided in Ref.~\cite{Zeller:2002he}, and eventually compute the total number of NC DIS $N_{\rm SM, tot}$. Hence, the number of DIS non-calculable events $N_{\rm SM, noncalc}$ is provided by the difference between $N_{\rm SM, tot}$ and the number of events computed in the calculable region with \eqref{eq:calcdisbkg}, $N_{\rm SM, nonpert}+N_{\rm SM, pert}$.

We note that the procedure described above is conservative, as it underestimates the size of the perturbative region. In particular, its lower boundary is set by the upper edge of the hatched area in \cref{fig:N}, which corresponds to the contour $Q^2 = 2m_p^2$ evaluated with the maximum neutrino energy $E_{\nu, \rm max}$.  This is more restrictive than using the lower edge of the solid red region, which corresponds to the same $Q^2$ contour with minimal energy available in the flux, $E_{\nu, \rm min}$. 

The test statistics used to derive the most conservative sensitivity for SHiP in Fig.~\ref{fig:moneyplot} is given by
\begin{equation}
\lambda = \frac{N_{\rm BSM,pert}^2}{N_{\rm SM,pert}}\,,
\end{equation}
where only events in the perturbative regime ($Q^2 \geq 2\,\mathrm{GeV}^2$) are included.
A better sensitivity is instead obtained including also the nonperturbative region with a modified test statistics 
\begin{equation}
\lambda = \frac{N_{\rm BSM,pert}^2}{N_{\rm SM,pert}} + \frac{N_{\rm BSM,nonpert}^2}{N_{\rm SM,nonpert} + \epsilon_{\rm rej} N_{\rm SM,noncalc} + \left(N_{\rm SM,nonpert} + \epsilon_{\rm rej} N_{\rm SM,noncalc} \right)^2}\,,
\end{equation}
where $\epsilon_{\rm rej}$ quantifies the efficiency of rejecting DIS events occurring in the noncalculable regime, relative to those from the calculable one. As shown in \cref{sec:Nscatt}, for fixed hadronic energy and angle in the laboratory frame, the exchanged momentum tends to be larger in BSM events than in SM ones, whereas the jet mass is typically larger in SM events. Lower invariant hadronic masses are generally associated with fewer reconstructed tracks, while higher values of $Q^2$ can lead to differences in jet substructure. A more refined analysis targeting these distinctive features in hadronic jet structure, together with more precise experimental determination of $\theta_{\rm j}$, $E_{\rm j}$ and $m_{\rm j}$, could help discriminate BSM events from SM backgrounds.

The strategy adopted at FASER-FPF resembles the one of NuTeV experiment where we impose a cut on the hadronic energy, requiring \( E_j > 20 \) GeV. Given the higher-energy neutrino flux at FASER-FPF and NuTeV compared to SHiP, this threshold ensures that the events lie safely within the perturbative regime.

Quasi-elastic scatterings — i.e., processes with $x \approx 1$ and $m_j^2 \approx m_p^2$ — are expected to account for approximately $10\%$ of the total number of SM neutral current scattering events at SHiP, while being negligible at FASER due to the higher center-of-mass energy (see Ref.~\cite{Formaggio:2012cpf}). However, such events are expected to be distinguishable from deep inelastic scattering ones, where the number of hadronic tracks in the final state is larger, due to the higher invariant hadronic mass $m_{\rm j}^2 > 2 m_p^2$. For this reason, we require for deep inelastic scattering events $m^2_{\rm j}>2m_p^2$ (see \cref{fig:N}).

\section{Neutrino Dipole interactions}\label{app:dipole}

In this section, we show an example of an operator which allows for UV dominated invisible jet production for $\Delta_N\geq 5/2$.
Consider the interaction term:
\begin{equation}\label{eq:composite_dipole}
    \Delta \mathcal{L} = \frac{H L \mathcal{O}_N \sigma^{\mu\nu} B_{\mu\nu}}{\Luv^{\Delta+1/2}} \,.
\end{equation}
After EW symmetry breaking this interaction can be seen as a ``composite dipole'' portal for the neutrino involving both the photon and $Z$ boson. Notice that the $Z$ exchange is suppressed by a factor of $\sin^2 \theta_W(\sqrt s/v)^4$ compared to photon exchange.
In the case of photon dipole, the operator mediates  neutrino scattering in a process very similar to the one considered in Fig.~\ref{fig:el}, without having an off-shell neutrino propagator.
Consider the process $f+\nu\to f+{\rm DS}$, where $f={e,q,\bar q}$: the squared matrix element averaged over the initial spin of $f$ and summed over the final spins and the DS states is
\begin{equation}
    \vert \bar M_\gamma\vert^2=\frac{2c^2_f e^2\cos\theta_W^2v^2A_N}{4\Lambda_{\rm UV}^{6}}\frac{-2su+m_N^2(s+u)}{t}\left[\frac{m_N^2}{\Lambda_{\rm UV}^2}\right]^{\Delta_N-5/2}\,,
\end{equation}
where $c_f$ is the fraction of electron charge carried by the fermion. The fully differential cross-section is
\begin{equation}
    d\sigma_\gamma=\frac{2c^2_f e^2\cos^2\theta_Wv^2A_N}{8s\Lambda_{\rm UV}^{6}}\frac{-2su+m_N^2(s+u)}{t}\left[\frac{m_N^2}{\Lambda_{\rm UV}^2}\right]^{\Delta_N-5/2}\frac{dm_N^2}{2\pi}\frac{dp_N^3}{2E_N(2\pi)^3}\frac{dp_4^3}{2E_4(2\pi)^3}\delta^4(p_1+p_2-p_N-p_4)(2\pi)^4\,.
\end{equation}
We deal with the phase space as in \cref{app:nucleon} obtaining the following expression
\begin{equation}
    \frac{d\sigma_\gamma}{dy dm_N^2}=\frac{c^2_f e^2\cos^2\theta_Wv^2A_N}{16\pi^2 s^2\Lambda_{\rm UV}^{6}}\frac{m_N^4+m_N^2s(y-2)+2s^2(1-y)}{y}\left[\frac{m_N^2}{\Lambda_{\rm UV}^2}\right]^{\Delta_N-5/2}
\end{equation}
Focusing on the case $f=e$, we can integrate over $y$ regularizing the IR contribution with a cutoff $E_{e,min}$ and considering the dominant term with the logarithmic divergence, we obtain
\begin{equation}\label{eq:dipole_diffxsec}
    \frac{d\sigma_\gamma}{dm_N^2}=\frac{\alpha_{\rm em}A_N\cos^2\theta_W}{4\pi}\frac{v^2}{\Luv^{6}}\left[2-\frac{m_N^2}{s}\left(2-\frac{m_N^2}{s}\right)\right]\log\frac{1-m_N^2/s}{E_{e,min}/E_\nu}\left[\frac{m_N^2}{\Luv^2}\right]^{\Delta_N-5/2} 
\end{equation}

which is UV dominated for $\Delta_N>5/2$ as expected.
Scattering bounds on this operator are weaker due to the higher dimensionality of the portal at fixed dimension of $\mathcal{O}_N$. 

Choosing $\Delta_N=5/2$, the number of signal events produced by $N_\nu$ neutrinos with mean energy $\bar{E}_\nu$ scattering over nucleons with target density $\sigma_{\rm target}^{-1}$ through the photon dipole portal is
\begin{equation}
    N_S\approx \sigma_{\rm target}^{-1}N_\nu\frac{\alpha_{\rm em}}{2\pi}\frac{v^2m_p \bar{E}_\nu}{\Luv^6}\sim10^3\times\left(\frac{N_\nu}{10^{15}}\right)\left(\frac{\sigma_{\rm target}^{-1}}{0.2\,{\rm GeV}^2}\right)\left(\frac{\bar{E}_\nu}{300\,{\rm GeV}}\right)\left(\frac{\Luv}{300\,{\rm GeV}}\right)^{-6}\,.
\end{equation}
By choosing $\Delta_N = 5/2$ in  \cref{eq:composite_dipole}, we get the following estimate for the lifetime for the decay to $\nu+\gamma$ inherited by the light dark states as a consequence of \cref{{eq:composite_dipole}}

\begin{equation}
    \tau \simeq 2\times 10 \,\mathrm{m} \left(\frac{10 \, \mathrm{MeV}}{\Lambda_\mathrm{IR}}\right)^5 \left(\frac{\Luv}{100 \, \mathrm{GeV}} \right)^6\,\ ,
\end{equation}
which is sensibly more short lived than the lifetime obtained from neutrino-portal interaction in Eq.~\eqref{eq:lifetime}.

Bounds on this composite dipole can be recasted following the approach of~\cite{Fernandez-Martinez:2023phj} for so called \emph{tensor currents}. These are essentially the operator of Eq.~\ref{eq:composite_dipole}, but replacing a $\mathcal{O}_N$ with an elementary fermion.
The main difference with respect to the elementary scenario is that given the steeper UV dominance of the composite dipole, scattering against nucleons is preferentially incoherent in the latter case, while for the elementary scenario it proceeds through coherent scatterings with low exchanged momenta~\cite{Magill:2018jla, Plestid:2020vqf}.
Bounds coming from upper-limit on cross-sections into invisible states, such as LEP mono-$\gamma$ or differential measurements of neutrino scatterings against electron/nucleons can roughly be rescaled from the bounds shown in Ref.~\cite{Fernandez-Martinez:2023phj}.
The best bound is LEP mono-$\gamma$ independently on the neutrino flavor and it excludes $\Luv \lesssim $ TeV for masses up to 100 MeV.
For higher masses, the boosted lifetime becomes comparable to the detector size. In this case we expect bounds coming from the produced states decaying back to SM particles. We expect stronger, but model dependent bounds in this case, and therefore will not further pursue this avenue. One should always keep in mind that with an analogous UV completion to the one discussed in Eq.~\ref{eq:uvmodel} the constraints coming from high energy physics processes could always be relaxed.

\bibliography{bibliography}
\end{document}